\documentclass[preprint,12pt,english]{elsarticle}

\usepackage{graphicx}
\usepackage{subfigure}
\usepackage{amssymb}
\usepackage{amsmath}
\usepackage{lineno}
\usepackage[section]{placeins}
\usepackage[T1]{fontenc}
\usepackage{float}

\pdfoutput=1

\begin{document}

\begin{frontmatter}



\title{Automated Calibration System for a High-Precision Measurement of
Neutrino Mixing Angle $\theta_{13}$ with the Daya Bay Antineutrino Detectors}


\author[2,1]{J.~Liu\corref{cor1}}
\author[1]{B.~Cai}
\author[1]{R.~Carr}
%\author[1]{R.~Cortez\fnref{fn1}}
\author[1,4]{D.~A.~Dwyer}
\author[2]{W.~Q.~Gu}
\author[2]{G.~S.~Li}
%\author[1]{J.~Pendlay}
\author[1,5]{X.~Qian}
\author[1,3]{R.~D.~McKeown}
\author[1]{R.~H.~M.~Tsang}
\author[3]{W. Wang}
\author[1]{F.~F.~Wu}
\author[1,5]{C.~Zhang}

\address[1]{Kellogg Radiation Laboratory, California Institute of Technology, 
Pasadena, California, USA}
\address[2]{Department of Physics, Shanghai Jiao Tong University, Shanghai, 
China}
\address[3]{Department of Physics, College of William and Mary, Williamsburg, Virginia, USA}
\address[4]{Lawrence Berkeley National Laboratory, Berkeley, California, USA}
\address[5]{Brookhaven National Laboratory, Upton, New York, USA}

%%\cortext[cor1]{jianglai.liu@sjtu.edu.cn}
%%\fntext[fn1]{Deceased}

\begin{abstract}
We describe the automated calibration system for the antineutrino
detectors in the Daya Bay Neutrino Experiment. This system 
consists of 24 identical units instrumented on 8 identical 20-ton 
liquid scintillator detectors. Each unit is a fully automated 
robotic system capable of deploying an LED and various 
radioactive sources into the detector along given vertical axes. 
Selected results from performance studies of the calibration system 
are reported.
\end{abstract}


\end{frontmatter}

%\linenumbers

\section{Introduction}

\label{sec:intro}
Transformations among three active neutrino flavors, 
also known as neutrino oscillations, 
have been firmly established by  
solar, atmospheric, long-baseline accelerator, 
and long-baseline reactor neutrino
experiments~\cite{PDG,MV}.
%\ %
%\footnote{all experimental data except LSND and the antineutrino data from MACRO
%BooNE%
%}\ 
Under the three-neutrino framework, 
mixing among neutrinos can be described by the so-called PMNS 
matrix \cite{Pon68,MNS}, a 3x3 unitary matrix which conventionally 
gets parametrized by three independent 
``Euler angles'' $\theta_{12}$ (solar mixing angle), $\theta_{23}$ (atmospheric 
mixing angle), and $\theta_{13}$, one complex CP phase $\delta$, 
and two complex Majorana phases if neutrinos are Majorana particles. 
$\theta_{23}$ ($\sim$45 deg) and 
$\theta_{12}$ ($\sim$ 30 deg) have been measured to 5\% accuracy~\cite{PDG}.
However, $\theta_{13}$ had remained elusive until a series of 
recent measurements
worldwide~\cite{DYB12,T2K11,MINOS11,DC12,RENO12}, particularly after 
a definitive measurement of 
electron-antineutrino disappearance at the Daya\ Bay reactors \cite{DYB12}. 

\begin{figure}[H]
\begin{centering}
\includegraphics[width=4.5in]{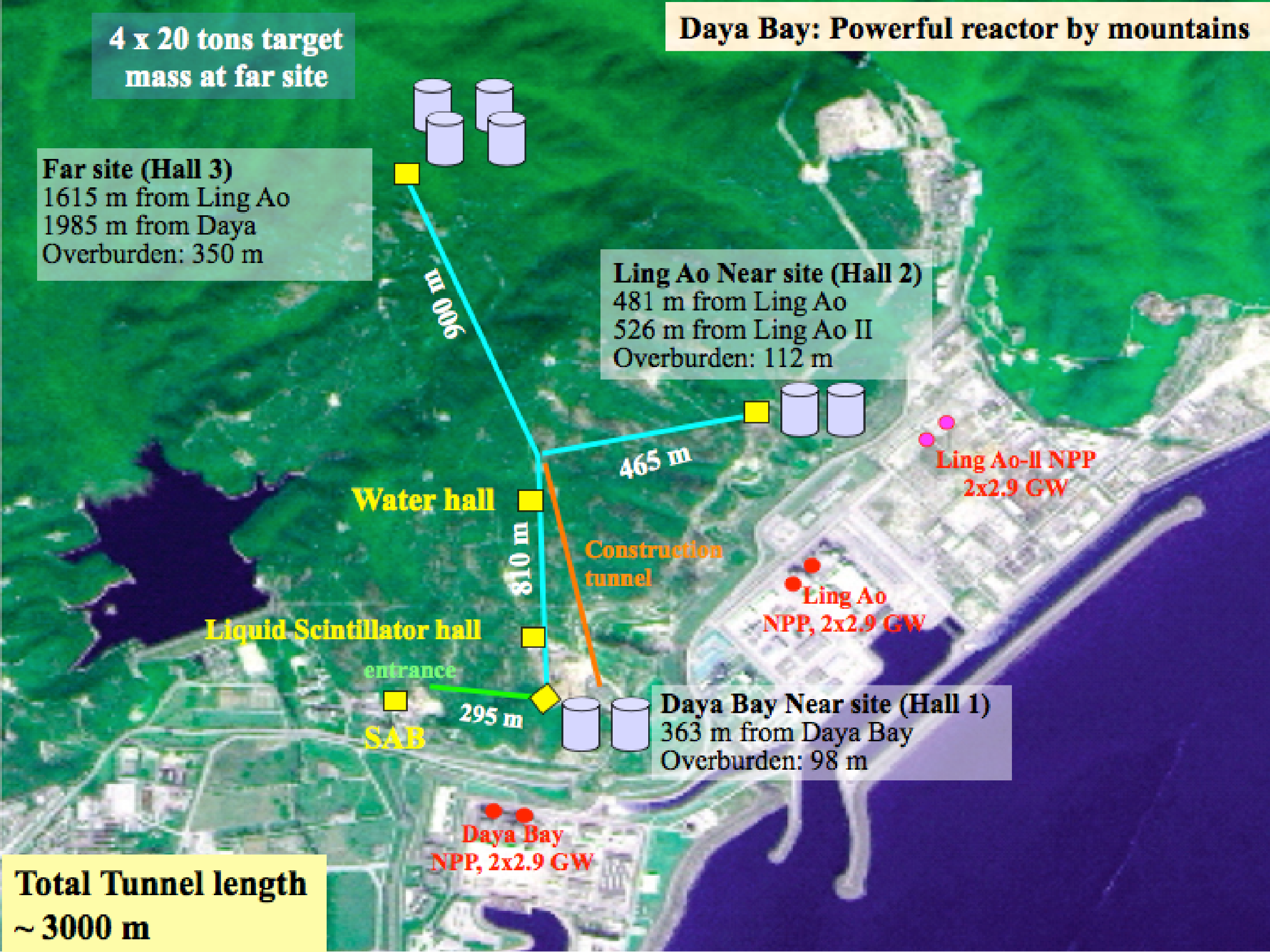}
\par\end{centering}

\centering{}\caption{\label{fig:The-layout-of-DayaBay}The layout of the Daya Bay experiment.}
\end{figure}

The Daya\ Bay Reactor Neutrino Experiment is located on the Daya
Bay reactor power plant campus in southern China. The power plant
currently hosts six 2.9\ GW$_{th}$ pressurized water reactors: the
Daya\ Bay complex has two reactors and the Ling\ Ao complex
has the other four. 
The layout of the Daya Bay experiment is shown in 
Fig.~\ref{fig:The-layout-of-DayaBay}.
The experiment has 8 ``identical'' antineutrino
detectors\ (AD) deployed in the three underground experimental halls:
two near halls to monitor the reactor antineutrino fluxes from the
Daya Bay and the Ling\ Ao complexes and one far hall to measure
combined flux around a baseline where maximal oscillation occurs. 
The near-far arrangement of the ADs can effectively cancel 
correlated uncertainties
in reactor antineutrino flux and detector efficiency, leading to 
high sensitivity to $\sin^{2}2\theta_{13}$\ \cite{Guo:2007ug}. 

A fully automated calibration system was designed and constructed to 
calibrate AD responses and to monitor their stability regularly. 
The rest of this article is organized as follows.
In Sec.~\ref{sec:requirements} of the paper, we
briefly describe the Daya Bay reactor neutrino experiment and 
design requirements of the calibration system. Design details will be 
given in Sec.~\ref{sec:design}, followed by a discussion of 
operational tests, quality control, and internal calibration 
of the system in Sec.~\ref{sec:mechanicalposition}. We will discuss 
calibration data taken 
during the commissioning tests, as well as 
in-situ calibration results during the physics running in 
Sec.~\ref{sec:dayaanalysis}, 
followed by a summary in Sec~\ref{sec:summary}. 

\section{Requirements of the calibration system}
\label{sec:requirements}
\subsection{Overview of experiment}
\begin{figure}[!htbp]
\begin{centering}
\includegraphics[width=4.5in]{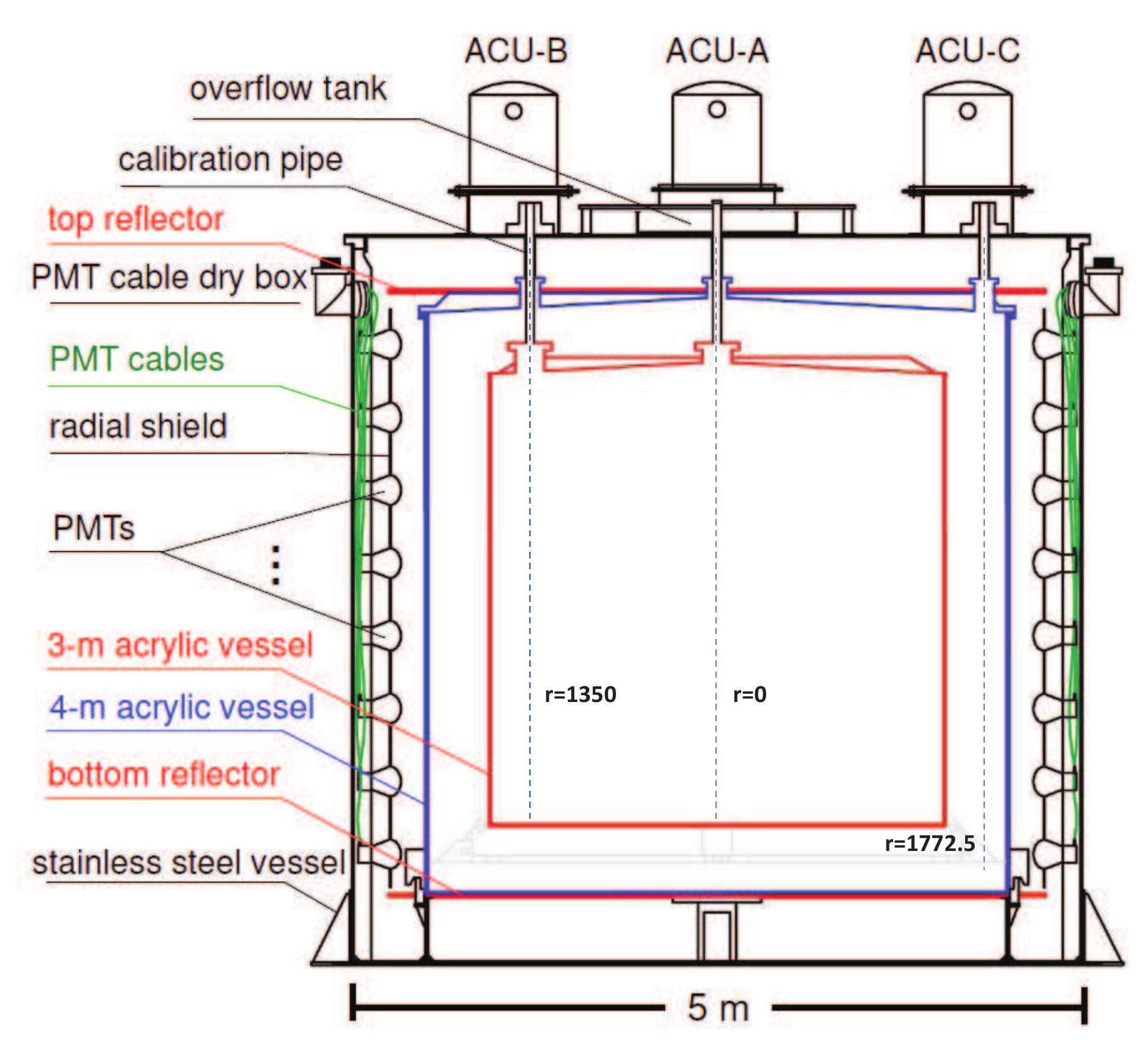}
\par\end{centering}
\caption{\label{fig:The-Daya-Bay-AD}
An illustration of the Daya Bay antineutrino detector structure. 
The inner acrylic vessel (IAV) holds the target, Gd-doped
liquid scintillator. The outer acrylic vessel (OAV) holds undoped 
liquid scintillator as the gamma catcher. The outermost zone inside the 
stainless steel tank where PMTs are located is filled with mineral oil. 
See text for more details. The target zone is
monitored by two ACUs, A (r=0 cm) and B (r=135.0 cm). 
ACU C (r=177.25 cm) monitors the gamma catcher zone. 
Three vertical source deployment axes are indicated by the dashed lines 
in the figure.}
\end{figure}

The primary goal of the Daya Bay reactor neutrino experiment 
is to make a sensitive measurement of $\sin^{2}2\theta_{13}$ 
to a precision <0.01 (90\% confidence level). Like other reactor 
neutrino experiments, electron-type antineutrinos are detected 
via the inverse beta decay~(IBD) process:
\begin{equation}
\bar{\nu_{e}}+p\rightarrow e^{+}+n\,.
\end{equation}
In this reaction, neutrino energy $E_{\nu}$ can be reconstructed by 
the kinetic energy of the positron $T_{e^+}$ as
\begin{equation}
E_{\nu} \simeq T_{e^+} + 1.8 \rm{MeV}\,.
\end{equation} 
Detailed discussion of the Daya Bay anti-neutrino detectors (AD) can be found 
elsewhere~\cite{Guo:2007ug,DYBNIM12}. For context, we reiterate a 
few key features here. Fig.~\ref{fig:The-Daya-Bay-AD} shows the design 
of the AD. 
The detector has three zones; the 
innermost acrylic vessel~(IAV) with 3-meter diameter holds the target
- 20t Gd doped LAB-based liquid scintillator. Final state positron 
in the IBD reaction annihilates with an electron after losing
its kinetic energy and releases two 0.511~MeV gammas, which we call the
\emph{prompt signal}. The final state neutron is first thermalized
and then gets captured either on Gd nuclei, emitting gammas with total energy 
of $\sim$8~MeV, or a proton, emitting a 2.2~MeV gamma.  
We call this the 
\emph{delayed signal}. The distinctive 8 MeV Gd capture 
signal defines the target mass zone reactions.
%\footnote{There is a spill-in effect which causes neutrons from 
%reactions outside the target zone get captures on Gd. 
%See Ref.~\cite{Guo:2007ug,An:2012pc}.%
%} 
%The strategy of the experiment is to select the IBD reactions in
%the target zone via their distinctive 8~MeV delayed signals. 
Several gammas are released from the neutron capture on Gd. Some gammas
would leak out without depositing energy in the target, causing a lower 
visible delayed signal energy. To reduce the gamma leakage, 
the second outer acrylic vessel (OAV), 4~m in diameter, is filled with 
undoped regular liquid scintillator, acting as the gamma catcher. To shield 
radioactive backgrounds from the environment and the
PMTs, mineral oil fills in the buffer between the 4~m vessel and
the 5~m diameter stainless steel detector tank. 

Despite the gamma catcher, not all delayed energy can be absorbed in the 
active volume containing Gd-loaded or plain scintillator, 
leading to a small leakage tail on the 
delayed energy spectrum. Daya Bay places a 6~MeV energy cut on the 
delayed signal. The prompt signal has at least 1.02~MeV annihilation 
energy deposited if the two 511 keV gammas do not escape the active 
volumes. The prompt signal energy cut is thus placed at 0.7~MeV considering
the energy resolution of the detector.

The measured neutrino spectrum for a given detector from a given 
reactor core can be written explicitly as
\begin{equation}
\label{eq:spec}
\frac{dN_{m}(E_{\nu})}{dE_{\nu}} = \frac{N_p}{4\pi L^2}\epsilon(E_{\nu})
\sigma_{\rm{IBD}}(E_{\nu})S_d(E_{\nu})P_{ee}(E_{\nu})\,,
\end{equation}
where $N_p$ is number of protons in the target, $L$ is the reactor 
core-detector distance (baseline), $\epsilon(E_{\nu})$ is the detector 
efficiency, 
$\sigma_{\rm{IBD}}(E_{\nu})$ is the inverse-beta decay cross section, and 
$S_d(E_{\nu})$ is the reactor neutrino flux from the core. $P_{ee}(E_{\nu})$,
the disappearance probability of electron antineutrinos, can be 
approximately expressed as
\begin{equation}
P_{ee} \sim 1 - \sin^2(2\theta_{13})\sin^2(\frac{1.267\Delta m_{31}^2 L }{E_{\nu}})
\end{equation}
with $L$ in meters and $E_\nu$ in MeV, and 
$\Delta m_{31}^2 \sim 2.35\times10^{-3}$~eV$^2$~\cite{PDG} being the absolute 
value of the mass square 
difference between ``1'' and ``3'' neutrino mass eigenstates. 
The extraction of oscillation parameters $\sin^2(2\theta_{13})$ and 
$\Delta m_{31}^2$ is essentially a fit of the measured spectrum 
$\frac{dN_{m}(E_{\nu})}{dE_{\nu}}$ to Eqn.~\ref{eq:spec} as a function of 
$E_{\nu}$. The calibration system, which we will elaborate in the rest 
of this article, is specifically designed to help us measure the energy 
for detected neutrinos.

The particle energy is converted nearly proportionally into light in the 
liquid scintillator, which then get detected by the photomultipliers in the AD. 
The calibration system of Daya Bay is designed to characterize detector
properties thereby to establish the conversion between 
measured PMT readout and the particle energy. Below is a summary 
of design guidelines.
\begin{enumerate}
\item Similar experiments in the past (Chooz and Palo Verde) both 
observed temporal change of liquid properties~\cite{Chooz03, PVxx}.
Therefore the Daya Bay calibration system needs to monitor detector performance 
on a regular basis, although we have performed bench measurements on the 
light attenuation in Gd-loaded liquid scintillator~\cite{Goett11}.

\item Key detector properties to be calibrated: PMT gains, timing, 
and photoelectrons (PE) to energy conversion in a range between 1 to 8 MeV. 

\item The Daya Bay detector is designed to be uniform in position 
response. To evaluate uniformity of the detector, 
calibration sources need to be loaded into different vertical and  
radial positions, including both Gd-loaded and regular liquid scintillator 
zones.

\item To minimize downtime, all detectors should be calibrated 
at the same time using the same procedure. Calibration needs to be 
simple and robust. In order to not introduce radon
background, a calibration should not require detector opening.

\item Cleanliness and chemical compatibility:
source assembly must not contaminate the liquid scintillator.

\item Radioactive background introduced by the calibration 
sources needs to be minimized during regular data taking.

\item The calibration system is the only moving detector component,
and reliability is a top consideration. Under no circumstance can it 
drop radioactive sources into the AD or damage the PMTs. Repair would 
be cumbersome and timing-consuming, and any need for them is not desirable.

\item Source deployment needs to be reproducible. The physical 
location of the source needs to be accurate to $<$5~mm, the 
level of position uncertainty of AD PMTs. 

\item The design of the electronics system should ensure that 
it does not introduce noise in the PMT readout.

\item Calibrations should be coordinated with 
data taking activities automatically, i.e. no start/stop runs by hand.
  
\end{enumerate}

A fully automated calibration system is the natural answer for requirement 1;
such a system has been used in other experiments~\cite{SNONIM}. 
Automation also reduces potential human-induced errors etc., as 
interlocks can be programmed in software. To satisfy 
Requirements 2-4, we designed three automated calibration
units (ACUs) to be mounted and sealed on the top of each detector permanently 
to allow for remote deployment of an LED (PMT gain and timing), 
a $^{68}$Ge source (1.022 MeV), and a combined source
of $^{241}$Am-$^{13}$C ($\sim$ 8 MeV) and $^{60}$Co (2.506 MeV) 
into the detector along three vertical axes 
(See Fig.~\ref{fig:The-Daya-Bay-AD}). 
Two of the ACUs (A and B) load sources along 
the center and an edge axis inside the target zone.
The third ACU (C) deploys sources along a vertical axis in the 
gamma catcher zone. Each ACU is an individual robotic system 
consisting of pulley/wheel and stepper motor assembly powered
by custom electronics, controlled by National Instruments~\cite{ref:NI} motion 
control, DAQ cards, and custom LabVIEW software. 

Other requirements from the list will be addressed in the remainder of this 
article. Requirements 5 and 6 will be addressed in Sec.~\ref{sub:mat-selection}
, as well as in Sec.~\ref{sec:turntable}. The insurance of
system reliability (Requirement 7) is discussed in the mechanical design 
(Sec.~\ref{sub:hardware-design}), software design 
(Sec.~\ref{sub:controlware}), as well as in quality control tests 
(Sec.~\ref{sec:qa_tests}).
The position accuracy of source deployment (Requirement 8) 
will be discussed in 
Sec.~\ref{sec:pos_calib}. Requirements 9 and 10 will be addressed in 
Sec.~\ref{sec:electronics} and \ref{sec:software_control}, respectively.

\section{Design of the automated calibration system}
\label{sec:design}
\subsection{Material selection}
\label{sub:mat-selection}
Materials used in the ACUs are selected based on two considerations: 
chemical compatibility with the LAB-based liquid scintillator, and 
the level of radioactivity.

For the chemical compatibility, 
we separate materials into two classes, those that will A) be immersed 
in or in contact with the liquid (during source deployment), and B) only be 
exposed to the vapor of liquid scintillator. After extensive chemical
assays at the Institute of High Energy Physics and the 
Brookhaven National Laboratory, we decided that 
the materials used in A) had to be or be enclosed in
either acrylic or Teflon (PTFE or FEP). For B), we relaxed the list to 
acrylic, Teflon, 316 stainless steel, Viton-A (Dupont), and ceramic. 
To bond acrylic parts, we used Weld-on 3 or 4 acrylic welds (IPS Corp.). 

We used an ORTEC HPGe detector at Caltech with lead and Cu shielding 
to carry out radioactivity assay of materials used in the ACU. 
Specs are set to 0.6 Bq/kg on $^{238}$U, and 0.4 Bq/kg on $^{232}$Th, 
2.6 Bq/kg on $^{40}$K based on detector Monte Carlo\footnote{To set the scale,
the AD single rate due to intrinsic radioactivity is about 70 Hz
with a threshold of $\sim$0.7~MeV.}.
All welding of flanges were performed with non-thoriated electrodes and 
ER308L welding rod (measured to be low radioactivity) where needed.

\subsection{Mechanical design}
\label{sub:hardware-design}
\begin{figure}[!htbp]
\begin{centering}
\includegraphics[width=5in]{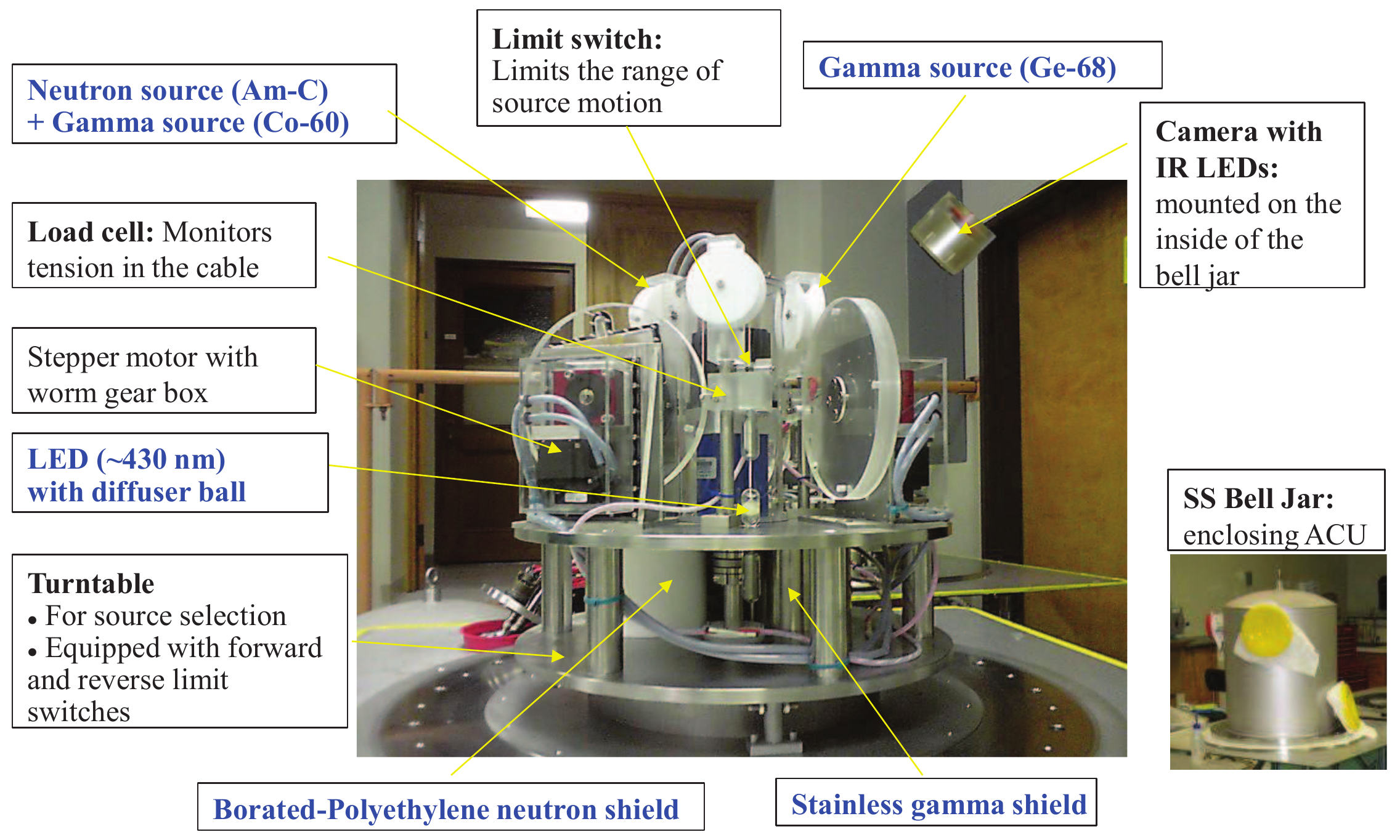}
\par\end{centering}
\caption{\label{fig:ACU-overview}An overview picture of an ACU.}
\end{figure}

\begin{figure}[!htbp]
\begin{centering}
\includegraphics[width=4in,angle=90]{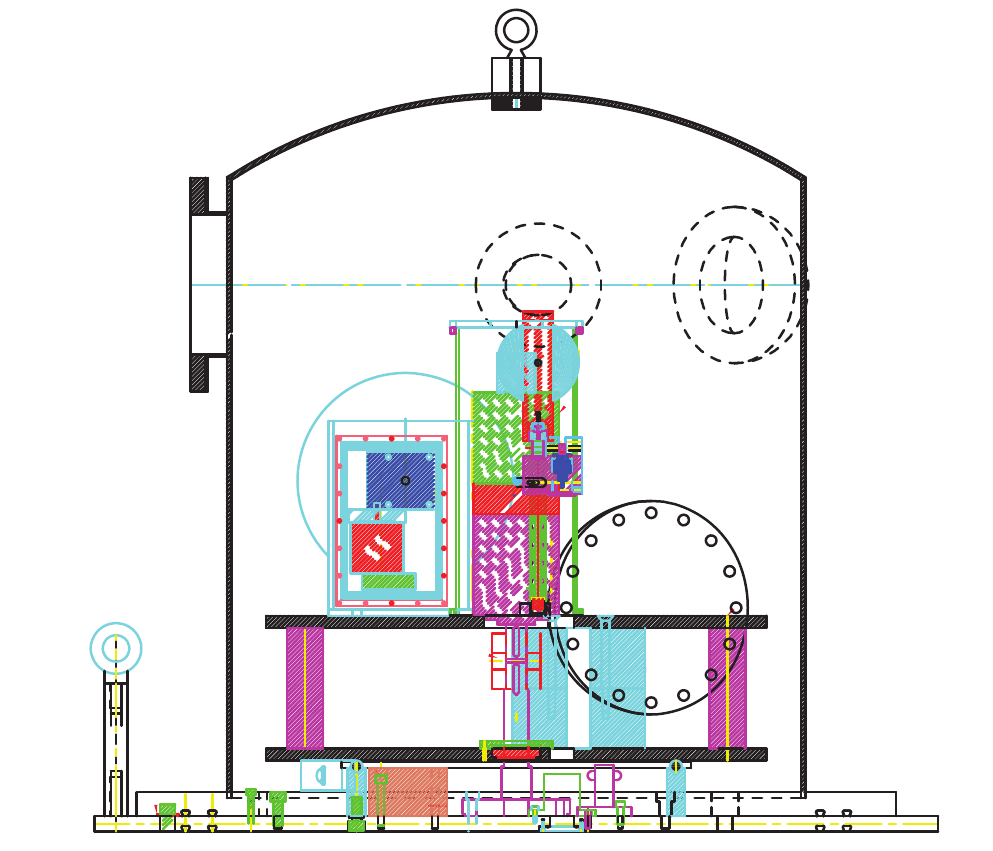}
\par\end{centering}
\caption{\label{fig:ACU-sideview}Sideview of an ACU.}
\end{figure}

% I have made changes to this section -- raymond
As shown earlier in Fig.~\ref{fig:The-Daya-Bay-AD}, 
each antineutrino detector is instrumented 
with three identical ACUs.
An overview image of an ACU and its major components is shown in 
Fig.~\ref{fig:ACU-overview}.
Each ACU is enclosed in a bell jar with 24-inch diameter and 30-inch 
height (Fig.~\ref{fig:ACU-sideview}). All ports on the bell jar are sealed 
with Viton-A o-rings.
The bottom plate (5/8-inch thick, 35-inch diameter) of the 
bell jar supports the interior structure of the ACU. 
The underside of this plate seals to 
a 24-inch interface flange on the SSV lid. The weight of each 
ACU is about 314 kg.
%The bell jar and the bottom plate sealed the calibration port from 
%water. 
%Employed between 
%the ACU bottom plate and the support spool on the AD lid is another 
%double O-ring which can be pumped out for leak checking and hence ensuring 
%seal quality. 
%The bottom plate supports the whole ACU and, at the same time, 
%acts as a reservoir for the liquid scintillator that adheres to and drips 
%from the source enclosures.
Access to the detector for each ACU is 
provided through a single port on the lid of the detector.

\subsubsection{Turntable}
\label{sec:turntable}
\begin{figure}[!htbp]
\begin{centering}
\includegraphics[width=4.5in]{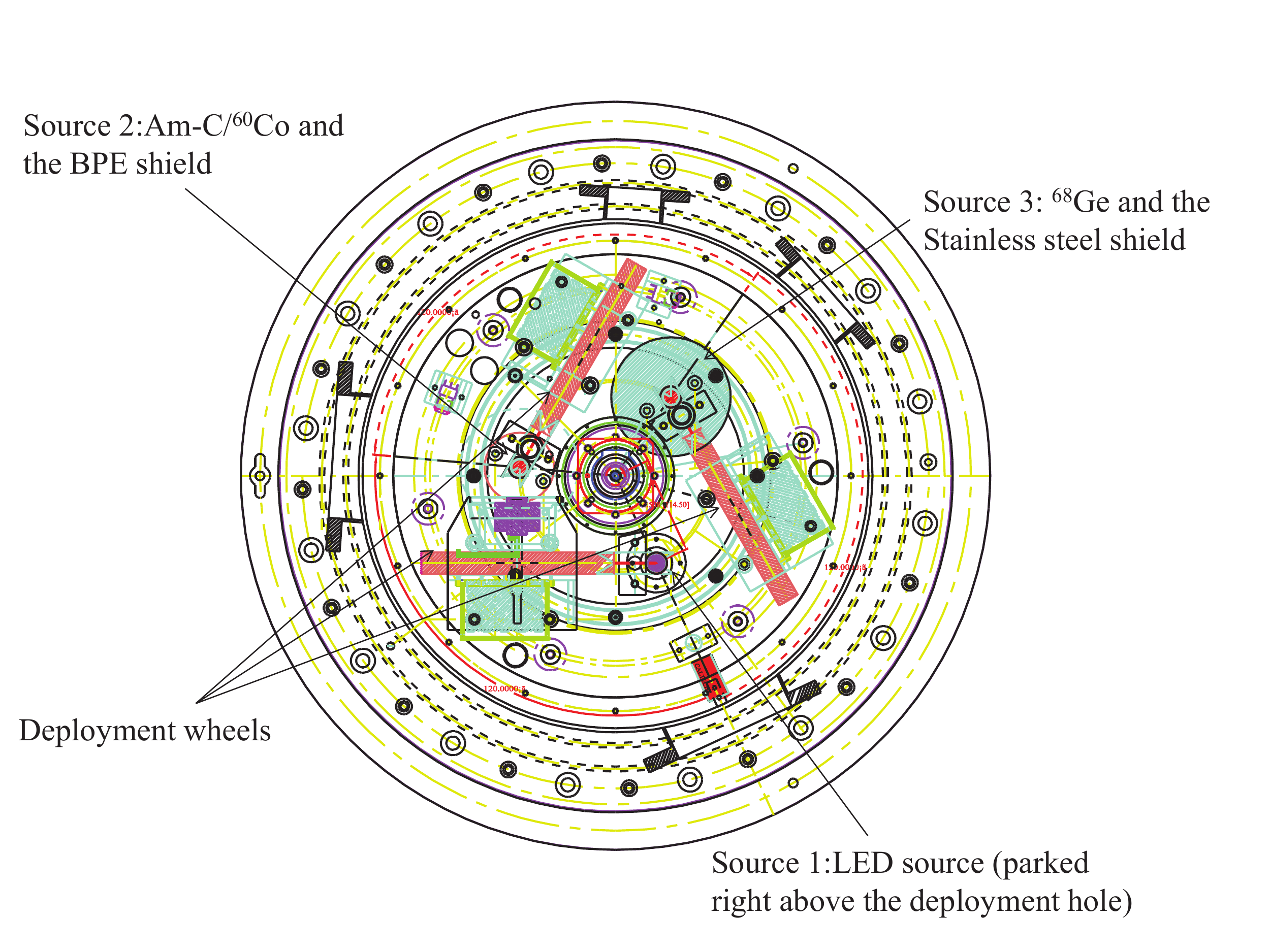}
\par\end{centering}
\caption{\label{fig:ACU-topview}Topview of an ACU.}
\end{figure}

The interior body of the ACU is a stainless steel turntable consisting 
of two plates (known as top and middle plates) and three mechanically 
independent motor/pulley/wheel assemblies (known 
as deployment axes) mounted on the top plate.
Each axis is capable of 
deploying a source (one of the radioactive sources or the LED) into 
the detector 
along the vertical axis (z-axis). The 5-inch separation between
the two plates hosts the storage housing for the radioactive sources.
The three axes are mounted with azimuthal
separation of 120 degree, as illustrated in Fig.~\ref{fig:ACU-topview}.

The turntable can turn clockwise or counter-clockwise by a 
stepper motor mounted at the center of the top plate. The center of 
each source is radially displaced from the center of the turntable 
by 114.5 mm. There is a 1-inch diameter hole at the same 
radial location on the bottom plate, aligning to a given penetration 
on the AD (see Fig.~\ref{fig:The-Daya-Bay-AD}), through 
which the source can get deployed into the detector volume.

\begin{figure}[!htbp]
\begin{centering}
\includegraphics[width=4in]{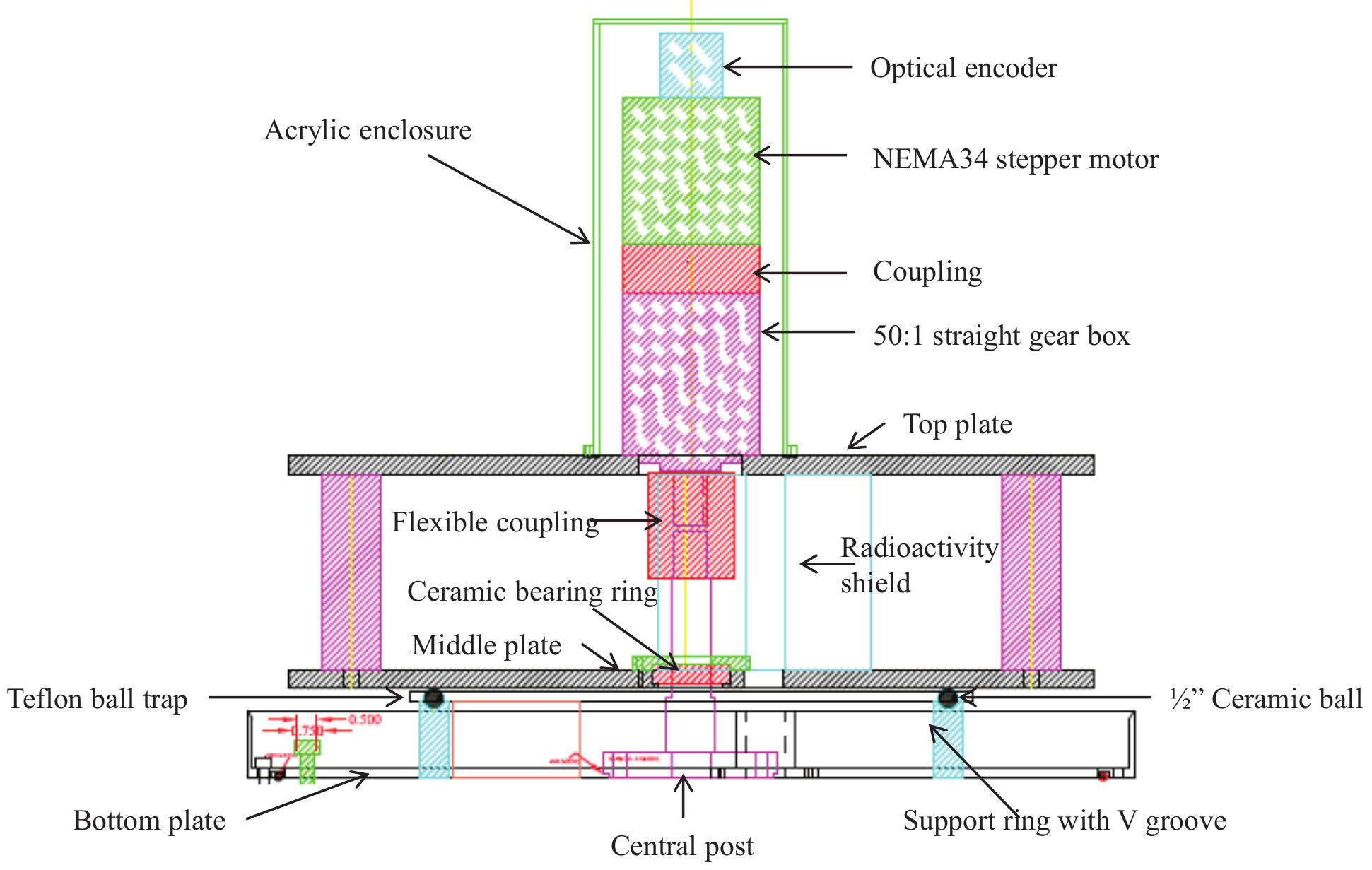}
\par\end{centering}
\caption{\label{fig:turntable}Drawing of a turntable.}
\end{figure}

Mechanical details of a turntable are shown in Fig.~\ref{fig:turntable}.
A fixed stainless steel post is mounted at the center of the bottom
plate, serving as the pivot of the rotation. A 2-inch high, 14-inch diameter 
stainless steel ring is mounted on the bottom plate.
The top surface of the ring is a V-groove, hosting 
eight 0.5-inch ceramic balls (evenly spaced by a Teflon spacer ring). This assembly
serves both as support to the turntable, as well as lubrication bearing 
to the turntable rotation. A NEMA34 stepper motor (Lin Engineering 
8718M-16DE-01) 
coupled to a 50:1 straight gear box is mounted at the center of 
the top plate. The shaft of the gear box couples to the center post
via a flexible bellow coupler (Wittenstein EC2 series).
The shaft of the stepper motor extrudes from the back, on which 
an optical encoder is attached to to digitize relative rotation 
between the motor shaft and its body. A ceramic bearing ring (Champion 
Bearings Inc.) is mounted between the center post and middle plate 
for alignment. To limit the range of the rotation,
a double-sided stop is mounted on the bottom plate.
Two pushbutton style stainless steel limit switches (Schurter Inc.) 
are mounted on the underside of the middle
plates to define the clockwise and counter-clockwise rotation limits (overall 
range 260 degrees).

Two radioactive sources in each ACU, when parked, are retracted into a
5-inch height, 2.25-inch thick cylinder (Borated polyethylene for
$^{241}$Am-$^{13}$C/$^{60}$Co and stainless steel for $^{68}$Ge) 
in between the top and middle plates to reduce background in the AD. 
A 2-inch thick, 3.25-inch diameter Borated polyethylene 
disk is located on the bottom plate right below the parked 
neutron source, serving
as additional neutron shielding. 

\subsubsection{Radioactive source deployment axis}
\begin{figure}[!htbp]
\centering
\subfigure[Drawing.] % caption for subfigure a
{
  \label{fig:deployment-axis-drawing}
  \includegraphics[height=5.5 in,angle=90]{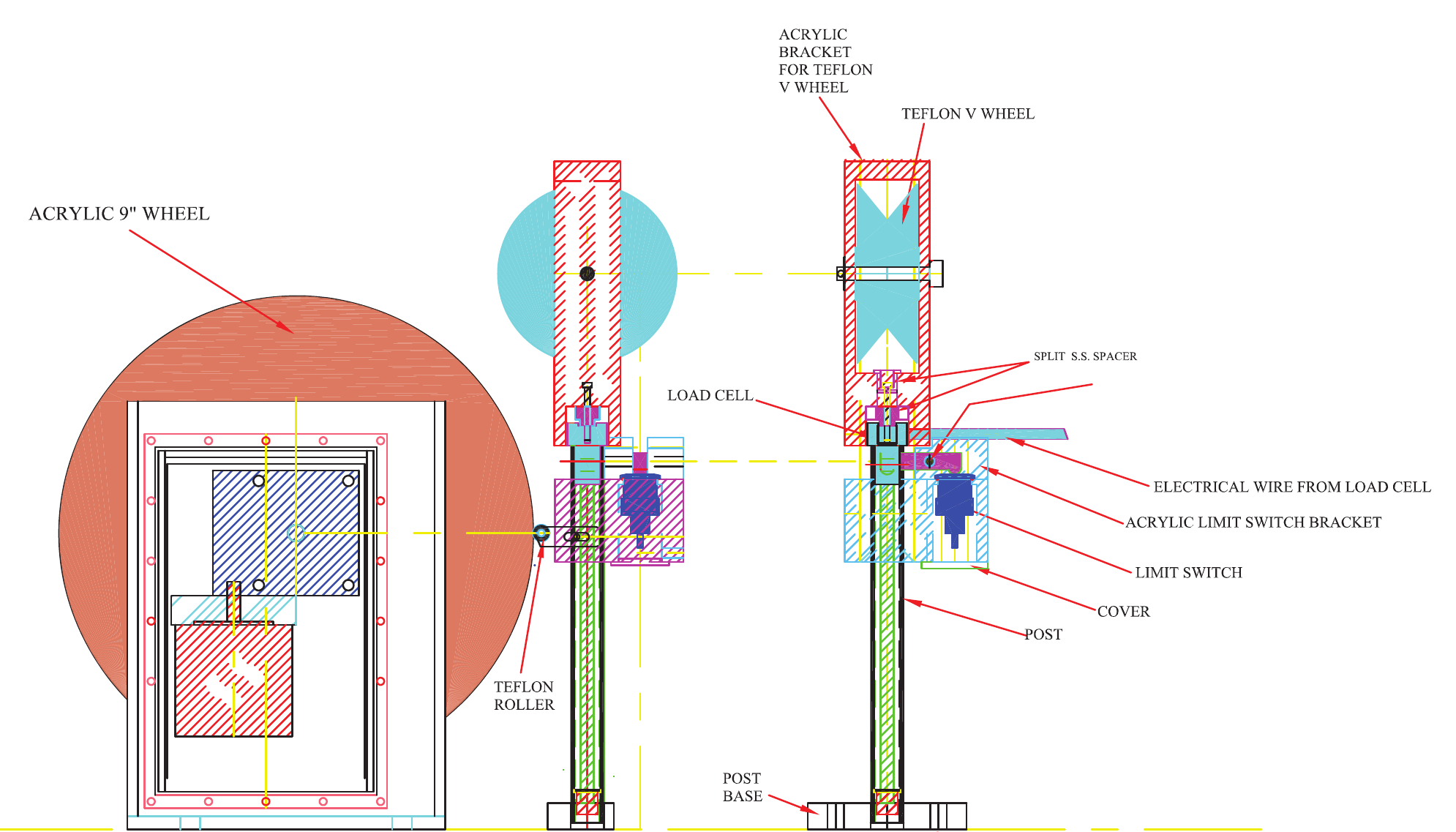}
}
\subfigure[Picture.] % caption for subfigure b
{
  \label{fig:deployment-axis-pic}
  \includegraphics[width=4 in]{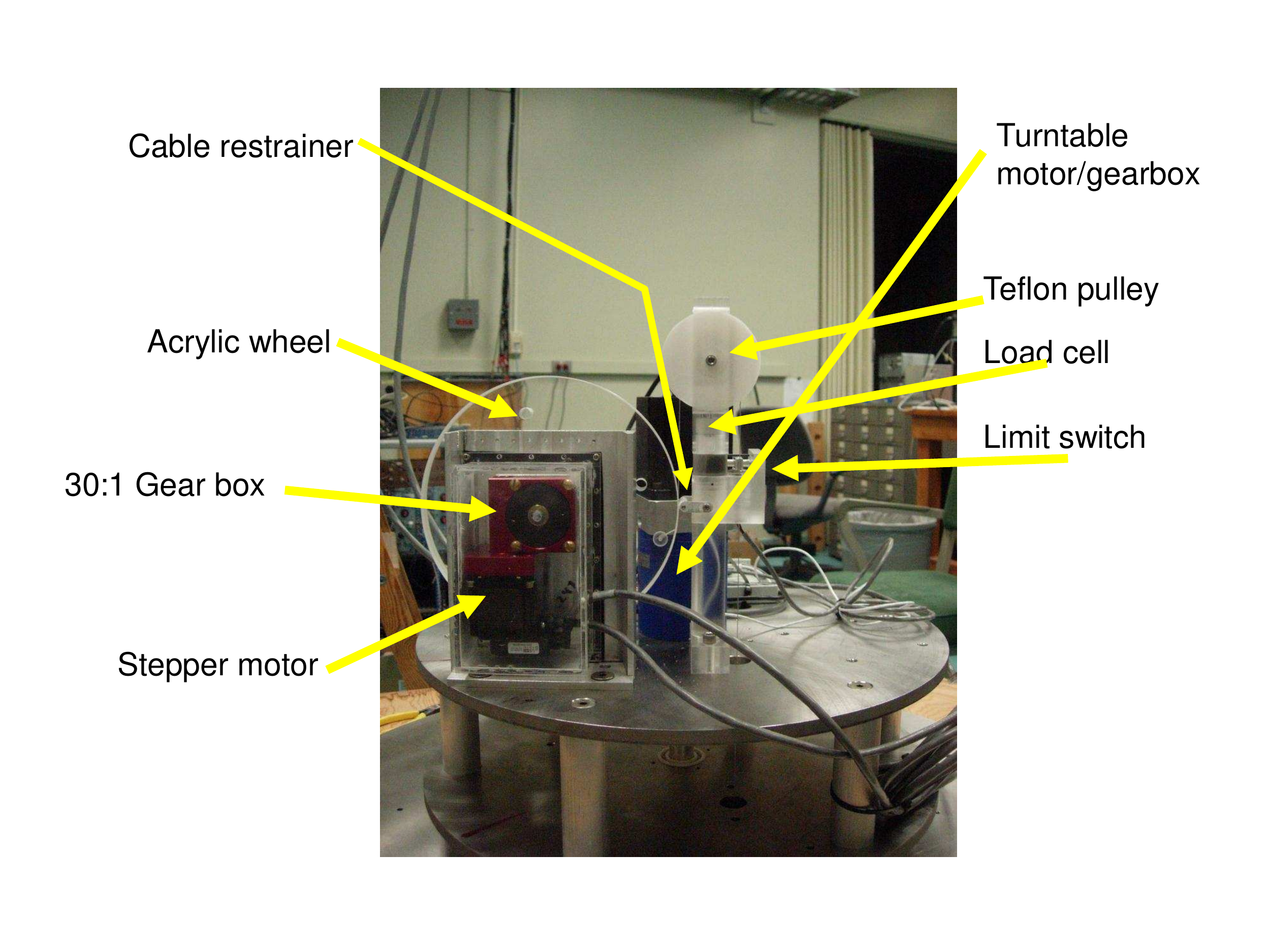}
}
\caption{       
 Drawing (a) and picture (b) of a single deployment axis.
}
\label{fig:deployment-axis} % caption for the whole figure
\end{figure}

Fig.~\ref{fig:deployment-axis} shows details of a radioactive 
source deployment axis. 
It consists of a NEMA 23 173 oz-in stepper motor (Lin Engineering Inc. 5718M), 
a 60:1 right-angle worm gear-box (Rino Mechanical Components Inc, 
PF30-60NM), a 9-in acrylic deployment wheel with helical grooves, 
and a Teflon pulley assembly. An optical encoder is mounted at the back of 
the stepper motor to encode rotations of the motor shaft. 

A flexible Teflon-jacketed stainless steel cable (McMaster Carr 3423T22, 
0.026-in OD), one end attached to the wheel, is wound in the helical groove
of the acrylic wheel. It then runs through a Teflon pulley assembly, and then 
attaches to the source assembly.
When the stepper motor drives the acrylic wheel to unwind the cable, the 
source is deployed into the detector. Gravity from the 
source assembly keeps the cable vertically straight. 

\subsubsection{LED source deployment axis}
\label{sec:LED_source_deploy_axis}
\begin{figure}[!htbp]
\begin{centering}
\includegraphics[width= 4.5 in]{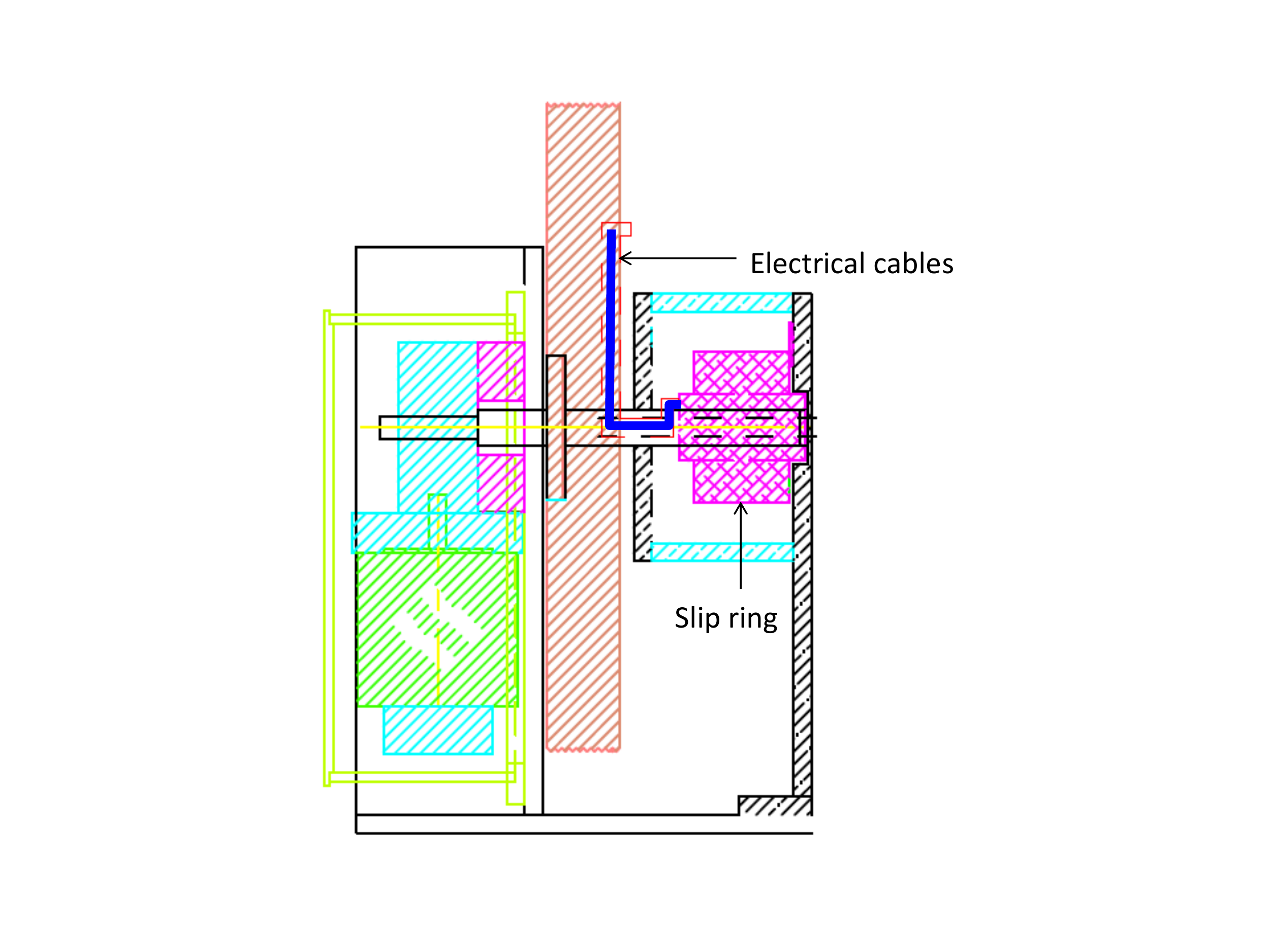}
\par\end{centering}
\caption{\label{fig:dep_axis_LED}Drawing of the LED source deployment 
axis (pulley assembly omitted).}
\end{figure}

The design of the LED deployment axis is shown in Fig.~\ref{fig:dep_axis_LED}.
Due to the need to transmit electrical signals to the LED, the design 
of the LED deployment axis differs from that of radioactive sources in a
few respects. It uses a miniature 
coaxial cable (Cooner Wire CW2040-3650 F, AWG 30 50 Ohm, 0.039-in OD) to 
both carry driving pulses, and serve as mechanical deployment cable. 
A LED driver (Sec.~\ref{sec:LED}) is mounted inside the 9-inch acrylic
wheel. A multi-wire electric slip ring (Moog AC6438-106) is mounted 
on a L-shaped bracket, which is fixed to the turntable. The rotary of the 
slip ring is attached to the shaft of the acrylic wheel. The slip ring carries
input electric signals from the turntable to the LED drivers inside the 
wheel.

\subsubsection{Failure protection}
\label{sec:failure_protection}
Numbers of protection features are implemented in the design of the 
source deployment system.
\begin{enumerate}
\item The gear box used in the driving unit is a worm gear which cannot be 
back-driven and is thus safe in case of power failure during a deployment. 

\item Each source assembly consists of two weights, one above and one 
below the source, by which the tension in the cable is 
maintained (Sec.~\ref{sec:source_assembly}). It is critical to monitor this 
tension to avoid the following:
a) tension that gets lost during the downward motion (bottoming out), 
causing the cable to unspool accidentally; b) source gets stuck 
during upward motion while the motor keeps turning and breaks 
the cable. A calibrated load cell (Honeywell Sensotec AL312-AT) is mounted 
below the Teflon pulley assembly. 
Source motion will be stopped immediately by the control software 
(Sec.~\ref{sub:controlware}) 
if the load cell value is outside the defined range. 
In case of bottoming out, the loss of tension due to the bottom weight 
signals the control system, while the top weight maintains sufficient 
tension in the cable. 

\item The acrylic shields for the upper/lower weights and the source are 
all made into round-headed bullet shape (Fig.~\ref{fig:source_assembly_pic}) 
to reduce the risk of the source getting stuck during upward motion.

\item A Teflon roller lightly presses against 
the acrylic wheel to retain the cable in the groove, while not
adding too much friction to the rotation of the wheel.
This further reduces the risk of accidental cable unspooling.

\item The turntable rotation is limited to a range of 260 
degree by two Schurter limit switches to avoid overtwisting 
of electronic cables.

\item All motions are made conservatively slowly. The sources move
at a speed of 7 mm/s, and the turntable rotates at a speed of 1.8 degree/s.

\item The upper limit of the motion is 
defined by the same Schurter limit switch as used on the turntable. 
When the upper weight hits a stop, it actuates the limit switch. 

\item The breaking strength for the stainless steel cable is 18 kg, and 
that for the coaxial cable is measured to be 2 kg. 
To further reduce the risk of accidental breakage of cables due to 
excessive tension, the torque of the stepper motor (via a current 
limiting resistor on the motor driver) is limited to provide 
a maximum tension of 1.0 kg in the cable. All cables (including the 
coaxial cable for the LEDs) 
are tested to this value. In case the source got stuck
and the load cell was malfunctioning, the stepper motor would ``slip'' 
(a mismatch between the stepping commander and the actual motion encoder)
due to insufficient torque, upon which the control software would 
kill the motion.

\item The turntable motor torque is limited the same way. In case a limit
switch was broken while ``stop'' was hit, the ``slip'' of the stepper motor 
would be detected by software.

\item The lower weight is made out of normal steel coated with Teflon, then 
enclosed in acrylic. In case of catastrophic cable breakage inside the AD, 
a rescue plan with a long rod and a permanent magnet could be developed. 

\end{enumerate}

\subsubsection{Auxiliary components}
\begin{figure}[!htbp]
\begin{centering}
\includegraphics[width=3 in]{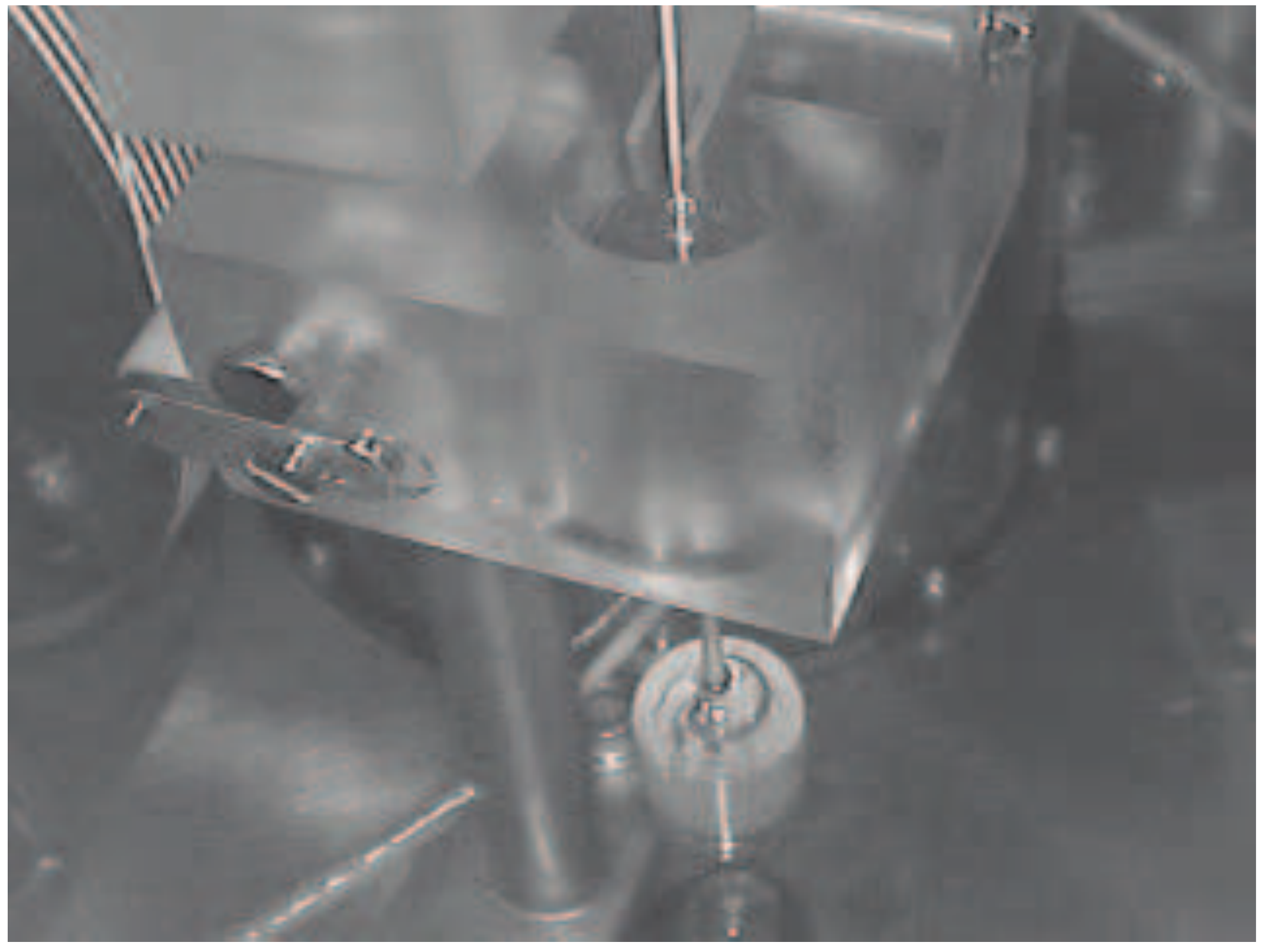}
\par\end{centering}
\caption{\label{fig:CCD} ACU CCD image during normal operation 
(source being stationary).}
\end{figure}

For monitoring purpose, a CCD camera (Q-See QOCCD) with infrared lighting 
is mounted on the inner wall of the bell jar. The infrared lights can be kept 
on during normal detector operation, and present no effects to the PMTs 
inside the detector. Fig.~\ref{fig:CCD} shows an image taken by the CCD camera.

\subsubsection{Materials}
The source assemblies are in direct contact with liquid scintillator. 
Both the coaxial cable for the LED axis and 
the stainless steel cable are Teflon-jacketed. The upper/lower weights and 
sources are all enclosed in acrylic. 

As shown in Fig.~\ref{fig:ACU-overview}, motor/gear reducer assemblies
for the turntable and source deployment axes are enclosed in acrylic 
housing. The CCD camera is also encapsulated in an acrylic shell. 
Teflon sleeves are used to cover all exposed non-Teflon 
electrical cables.

\subsubsection{Flange interfaces, dry pipe, and leak checking}
\label{sec:dry_pipe_leak_check}
The ADs and ACUs, during normal operation, are immersed in ultrapure water
in a water pool (an active muon veto detector). 
To be chemically compatible with ultra-pure water, 
all water seals are made with Viton-A o-rings. Standard ISO flange
seals are adopted wherever applicable. As a requirement, after the 
final bolt tightening, 
all seals have to be separately checkable to a leak rate with an upper limit 
of $1.5\times10^{-3}$ cc/s/atm for gas. Taking into account the fact that
the viscosity of water is about a factor of 50 larger than that of common 
gases, and that the ACUs are located 2 m under water (0.2 atm), 
this limit translates to a water leak about 200 g/year, which can be easily 
taken away by evaporation under dry nitrogen environment. 

Two types of leak checking techniques are employed in the ACU system: 
vacuum method
(pumping out a volume) and pressure/sniffing using Freon and 
commercial sniffer. Despite the outgassing
being an intrinsic background, the vacuum method generally has better 
sensitivity. The Freon sniffers, on the other hand, can also detect 
leak rate at $1.5\times10^{-3}$ cc/s/atm, based on tests with 
calibrated leaks.

\begin{figure}[!htbp]
\begin{centering}
\includegraphics[width=4 in]{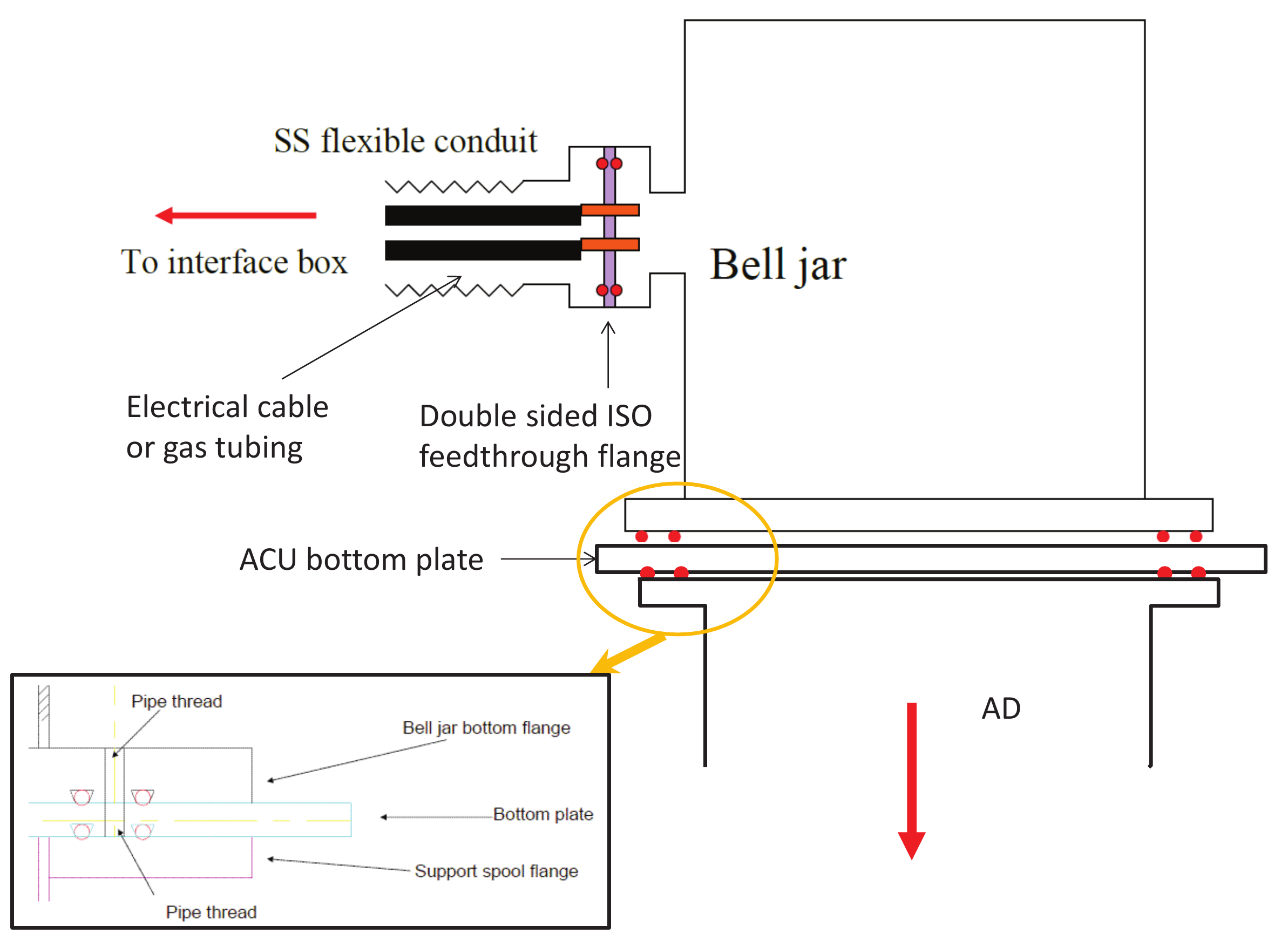}
\par\end{centering}
\caption{\label{fig:ACU_seal} Illustration of seals on the ACU, with 
o-rings indicated by red circles.}
\end{figure}

\begin{figure}[!htbp]
\begin{centering}
\includegraphics[width=4 in]{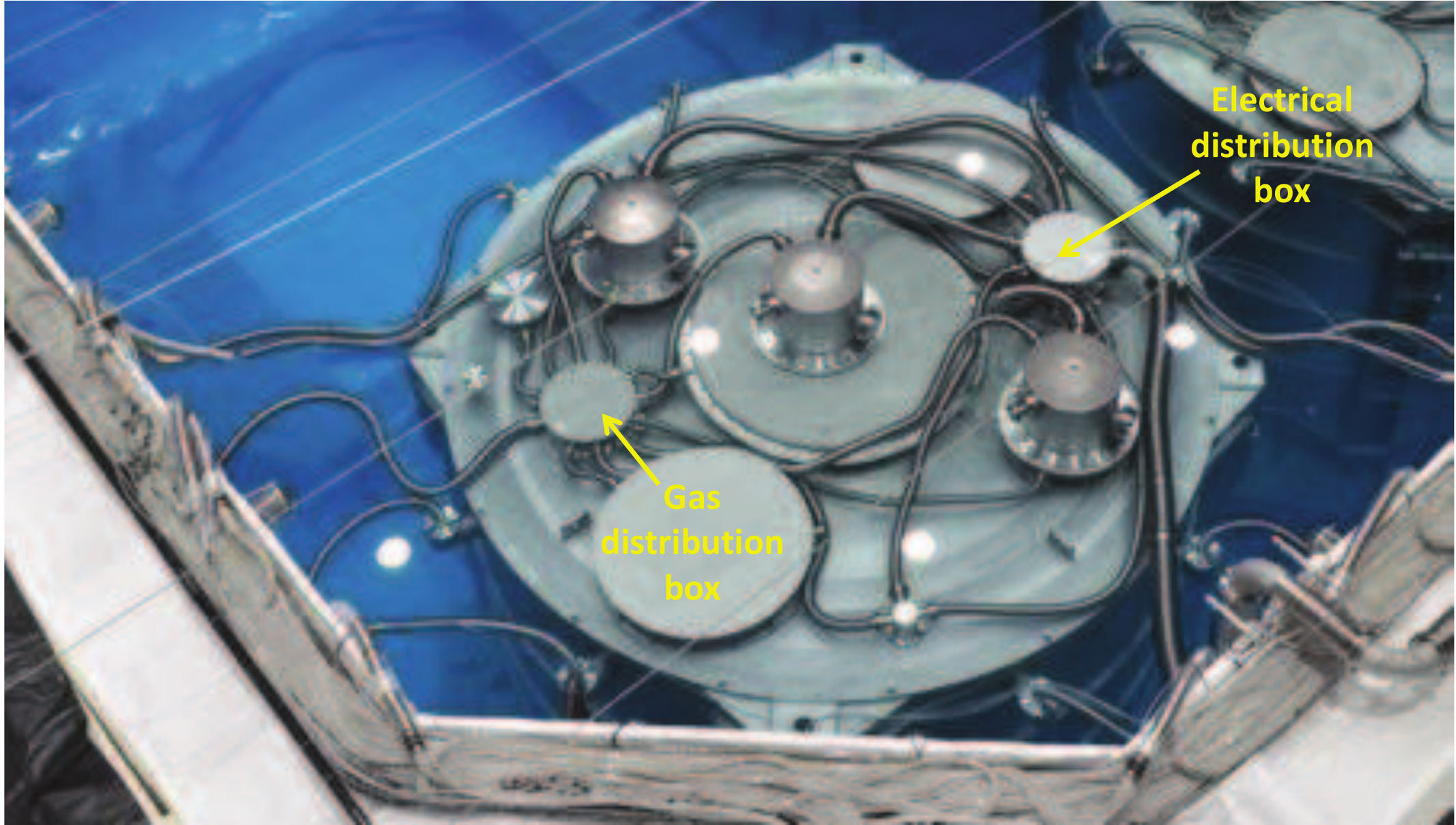}
\par\end{centering}
\caption{\label{fig:dry_pipe} A picture showing the ACUs and dry pipe 
system on the AD lid.}
\end{figure}

Flange interfaces on ACUs are illustrated in Fig.~\ref{fig:ACU_seal}. 
The ISO flanges
welded on the wall are sealed by double-sided electrical or gas feedthrough 
flanges. Outside the ACUs, the cables/gas lines run inside a so-called dry
pipe system, shown in Fig.~\ref{fig:dry_pipe}. 
The outer side of the feedthrough 
flange (facing away from the ACU) connects to a stainless steel 
flexible hoses (ISO63 or 100), which 
then connects to a stainless steel manifold 
chamber (known as the electrical or gas distribution box) where all cables 
or gas lines are consolidated. A stainless steel conduit 
assembly then takes all the cables and gas tubing out of the water pool.
The double-sided vacuum feedthrough flanges allow the ACU and 
dry pipe system be leak checked separately. In addition, the AD 
would be protected by these vacuum feedthroughs even if a water 
leak was developed in the dry pipe system. 

Details of seals below the ACU bell jar are shown in the inset of 
Fig.~\ref{fig:ACU_seal}.
The ACU bell jar seals to its bottom plate via a 
double o-ring seal, which then seals to the AD via another 
double o-ring seal. 
A 1/8-inch pumpout port is added between 
o-rings so that seals can be checked by pumping out the volume 
between o-rings. 

All ACU side flange seals 
were leak checked by pressurizing each ACU to 1.25 atm (absolute) with Freon 
{\underline{before the ACU gets installed onto the AD}}. 
After the ACU and cable installation, the entire dry pipe system 
is leak checked with 1.25 atm Freon. In this case, the open end 
of the dry pipe above water is sealed by a custom rubber seal 
designed at the Physical Science Laboratory at the University of Wisconsin 
with cables in situ.
%, as shown in Fig.~\ref{fig:dry_pipe_seal}. 
As a final check, 
the dry pipe is pressurized with 0.1 atm of nitrogen 
during the water filling when we looked for visible bubbles. 

\subsection{Calibration sources}
\label{sub:sourcedesign}
%The response of the detectors may have small difference, and it can
%lead to distortions of the measured anti-neutrino energy spectrum.
%It is important to thoroughly understand the detector properties before
%data taking and monitor the detectors performance and stability during
%the experiment. 
To track and study detector response, multiple calibration 
sources are implemented in each ACU. 
Three deployment axes in each ACU can deploy, {\underline{one at a time}}, 
a LED, $^{241}$Am-$^{13}$C/$^{60}$Co, or $^{68}$Ge, into the detector. 
They emit light either at a given time with controllable 
intensity (LED), or at given gamma energies: 1.022 MeV ($^{68}$Ge),
2.506 MeV ($^{60}$Co), 8 MeV ($^{241}$Am-$^{13}$C).

%To monitor and study detector response, calibration sources are 
%designed to emit lights with a fixed energy

%Calibration sources are designed emit
%to mimic inverse beta decay products in a controllable way. 
%Therefore, calibration sources must be deployed to
%the active volumes in the detectors at a routine basis, which mimic
%the inverse beta decay products and can be used to monitor the detector
%response to gammas from positron annihilation, neutron capture or
%other background. There are three deployment axes in each ACU: 
%the LED, $^{241}$Am-$^{13}$C/$^{60}$Co, and $^{68}$Ge. 

\subsubsection{LED}
\label{sec:LED}
Deployment axis 1 is designed to deploy a blue LED (Brite-LED BL-LUVB5N15C) 
with a 435 nm maximum wavelength. To make the light emission more 
diffuse, the LED is pocketed in a $\frac{3}{4}$ inch nylon ball. A miniature
coaxial cable is soldered onto the LED and they get epoxied
together with the nylon ball as shown in Fig.~\ref{fig:led}. 
The ball is then encapsulated in an acrylic enclosure with the cable 
going through the small hole
on the top of the acrylic enclosure (see Fig. $\,\ref{fig:led}$).
The cable is wound into the grooves of one of the acrylic wheels.
\begin{center}
\begin{figure}[!htbp]
\begin{centering}
\includegraphics[width=4.0 in]{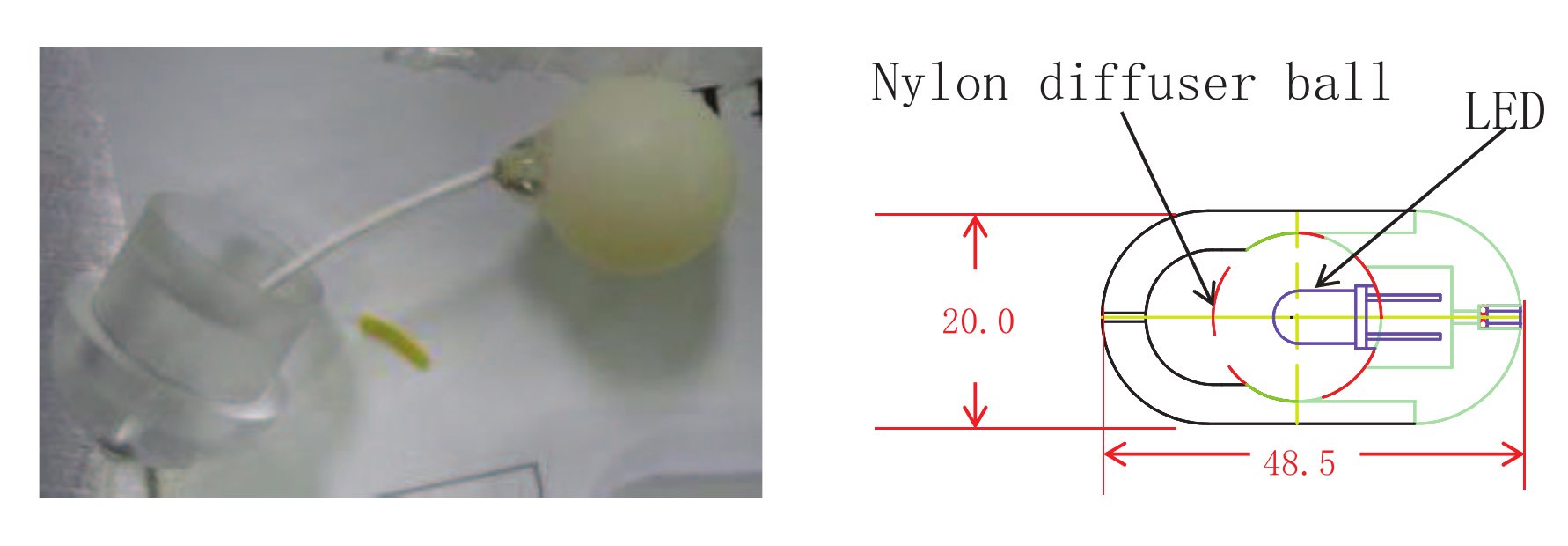}
\par\end{centering}
\caption{\label{fig:led} Left: picture of LED potted in the nylon ball; 
right: diagram showing the acrylic enclosure for the LED. }
\end{figure}
\par\end{center}

\begin{figure}[!htbp]
\centering
\includegraphics[width=4in]{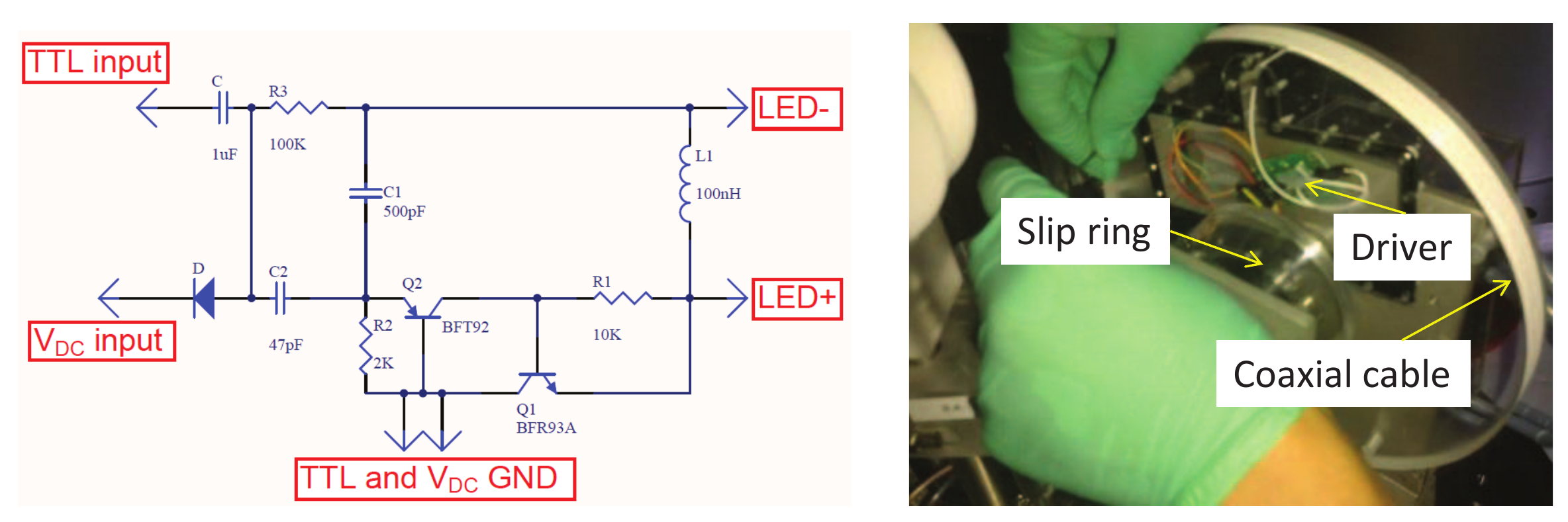}
\caption{\label{fig:LEDdriver} Left: schematics of the LED driver; right: picture of 
LED driver in the deployment wheel.
}
% caption for the whole figure
\end{figure}

The driver of the LED follows the design in~\cite{ref:LEDPulser},
also shown in Fig.~\ref{fig:LEDdriver}. The light flash is triggered by a fast 
TTL pulse, with its intensity controlled by a negative 
DC voltage V$_{DC}$. The driver board is kept in a pocket inside 
the 9-inch acrylic wheel, with its output
connected to the LED through the 5-meter long coaxial cable. 
An 100 nH inductor is in parallel with the coaxial cable on the driver.
This inductor is essential to give a narrow ($\sim$10 ns) light pulse.  
As discussed in Sec.~\ref{sec:LED_source_deploy_axis}, 
the electrical signals get transmitted onto 
the driver board via the slip ring joint.

\subsubsection{$^{137}$Cs scintillator ball}
\label{sec:scint_ball}
During detector commissioning before liquid scintillator filling (``dry run''),
it was desirable to use stable light source (LED light output intensity 
was found to be unstable at different locations) to scan the detector. 
We designed a special source made with $^{137}$Cs deposited at 
the center of 
a spherical scintillator. $^{137}$Cs primarily $\beta$-decays into $^{137m}$Ba 
($T_{1/2}=2.552$~m), which has about a 10\% probability of emitting K (624 keV) 
or L (656 keV) shell conversion electrons. The scintillation lights 
produced by these conversion electrons on the scintillator can be used 
as standard ``candles''.  

\begin{figure}[!htbp]
\centering
\includegraphics[width=4in]{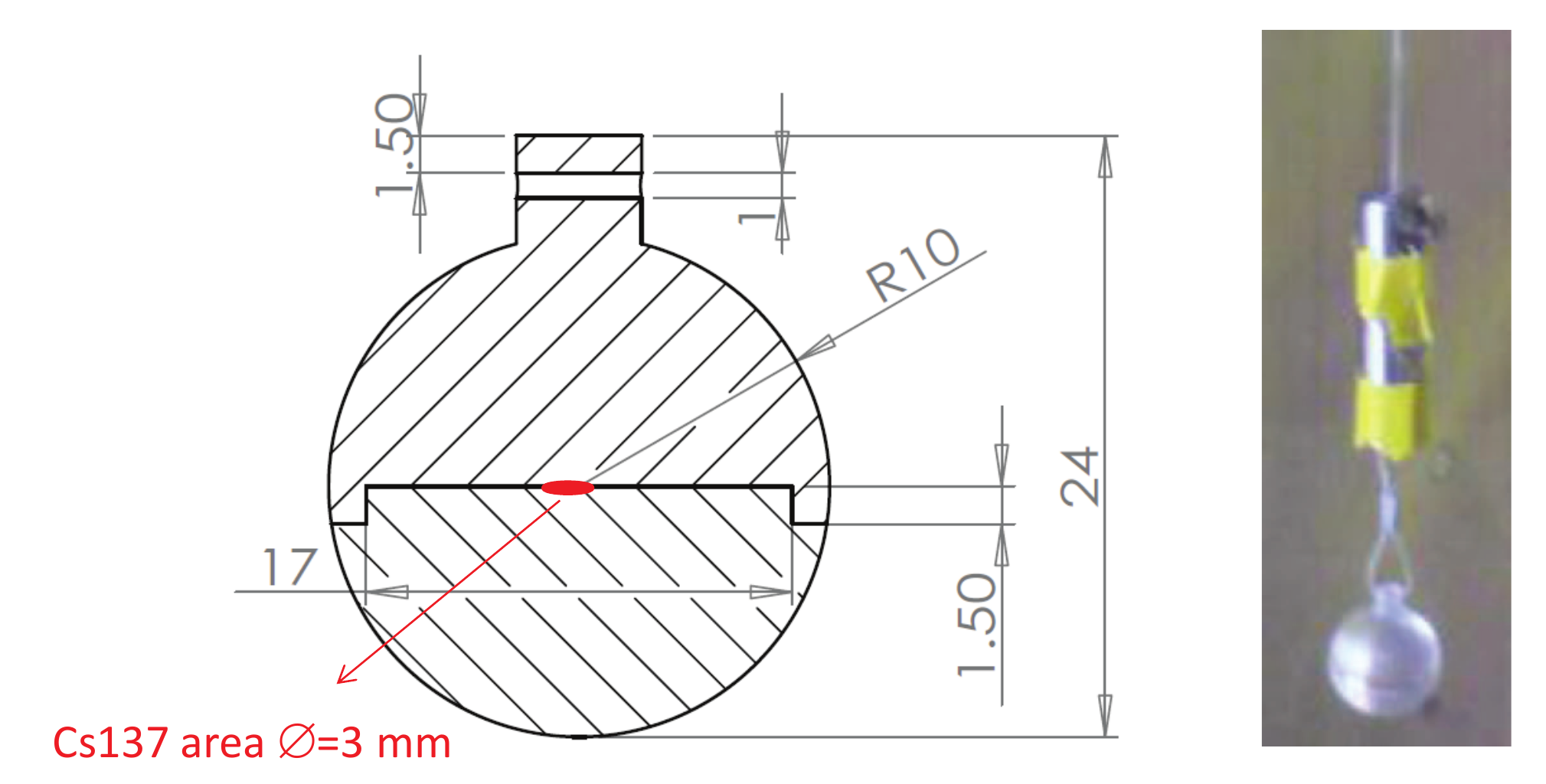}
\caption{\label{fig:scint_ball} Left: design of the scintillator ball; 
right: real picture taken during dry run.
}
% caption for the whole figure
\end{figure}

The design of the source is shown in Fig.~\ref{fig:scint_ball}, 
which was fabricated at 
the China Institute of Atomic Energy with 900 Bq of $^{137}$Cs deposited.  
A picture taken when the source was being prepared for the dry run is 
also shown in Fig.~\ref{fig:scint_ball}.

\subsubsection{Radioactive sources}
\begin{figure}[!htbp]
\begin{centering}
\includegraphics[width=4in]{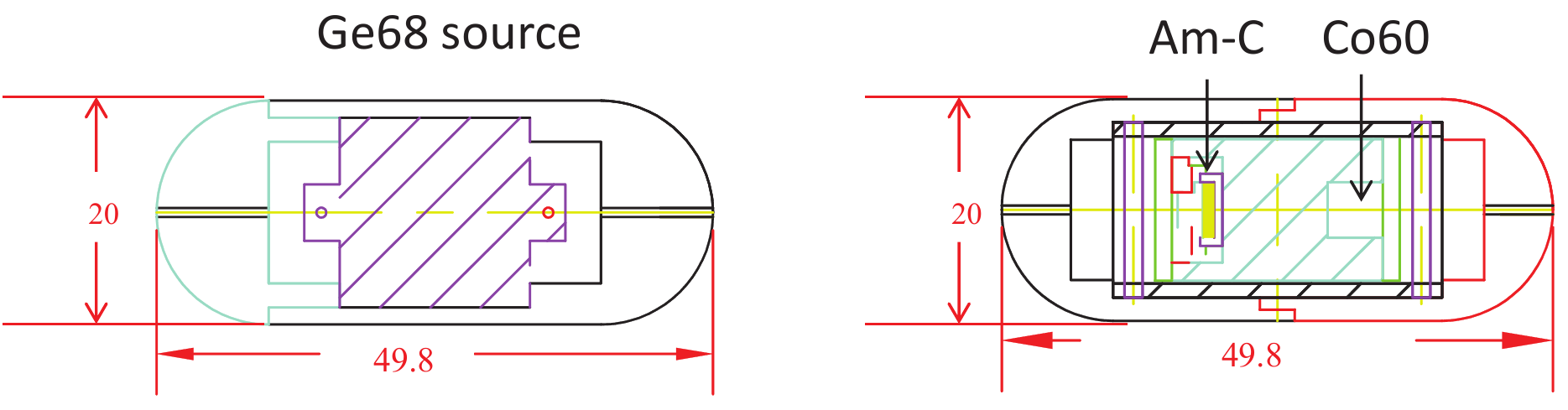}
\par\end{centering}
\caption{\label{fig:gamma-source}Acrylic enclosures for Ge68 (left) and 
AmC/Co60 (right).}
\end{figure}

Deployment axes 2 and 3 host radioactive sources. The radioactive
sources used in the calibrations are listed in 
Table \ref{tab:Radioactive-sources}.
\begin{table}[!htbp]
\begin{tabular}{|c|p{3in}|}
\hline 
Sources & Calibrations\tabularnewline
\hline 
Gamma source:$^{60}$Co & Energy linearity, stability, resolution, spatial and temporal variations, \tabularnewline
 & and quenching effect. Gamma line energies: 1.173+1.332 MeV\tabularnewline
\hline 
Neutron source:$^{241}$Am-$^{13}$C & Gd neutron capture energy: $\sim$8MeV total gamma energy\tabularnewline
\hline 
Positron source: $^{68}$Ge & Energy scale, trigger threshold, and PMT QE(relative). Energy: 0.511+0.511
MeV\tabularnewline
\hline 
\end{tabular}
\caption{\label{tab:Radioactive-sources}
Radioactive sources used in the ACUs.}
\end{table}
The energy of the sources is in a range from 1 MeV to 8 MeV. Axis
2 is for deploying a combination of $^{60}$Co (gamma) and $^{241}$Am-$^{13}$C
neutron sources (see Fig. \ref{fig:gamma-source}). The $^{60}$Co
source emits two gammas (1.173, 1.332 MeV) at about 100 Hz. 
Details of the $^{241}$Am-$^{13}$C neutron source will be discussed 
elsewhere~\cite{ref:AmC}. In brief, 
$\alpha$-particles from a 28~$\mu$Ci $^{241}$Am disk source 
impinge on a disk of compressed $^{13}$C powder and emit 
neutrons ($\sim$0.7 Hz) via $\alpha + ^{13}\rm{C} \rightarrow \rm{n} + ^{16}\rm{O}$. 
The neutron detection
characteristics of the detectors can be thoroughly calibrated by the AmC source.
The energy of $\alpha$s is attenuated by a gold foil (1~$\mu$m) to 
eliminate the production of excited states of $^{16}$O, thereby eliminating 
IBD-like background caused by correlated n-$\gamma$ emission. The 
residual backgrounds from $^{241}$Am-$^{13}$C are evaluated to be 
0.2\% compared to the IBD signals at the far site~\cite{DYB12}.

Axis 3 is responsible for deploying a $^{68}$Ge source ($\sim$10 Hz).
$^{68}$Ge decays into $^{68}$Ga with a $T_{1/2}$ of 271 days via electron capture. 
$^{68}$Ga then $\beta^{+}$-decay into $^{68}$Zn ($T_{1/2}=$67.7 minutes). 
Although $^{68}$Ga is a positron emitter, the majority of the positrons
emitted would annihilate in the source enclosure, making it effectively
a gamma source with two 511 keV gammas. The positron detection 
threshold of the detectors can then be calibrated.

\subsubsection{Source assembly}
\label{sec:source_assembly}
\begin{figure}[!htbp]
\begin{centering}
\includegraphics[width=3 in]{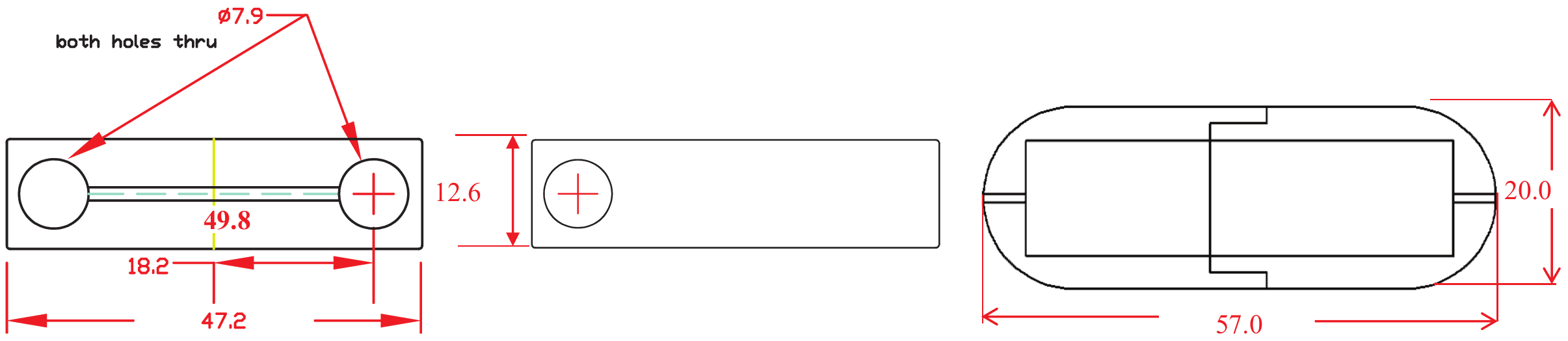}
\par\end{centering}
\caption{\label{fig:weight_shield}Top weight (left), bottom weight (middle), 
and their acrylic enclosures.}
\end{figure}

As mentioned briefly in Sec.~\ref{sec:failure_protection}, to maintain 
sufficient tension in the deployment string to avoid accidental unspooling, 
each source is accompanied with two weights, one above and one below the 
source. Dimensions of the weight and its acrylic enclosure are shown in 
Fig.~\ref{fig:weight_shield}. Weld-on 3 acrylic cement from IPS was used
to glue the enclosures.

\begin{figure}[!htbp]
\begin{centering}
\includegraphics[width=4in]{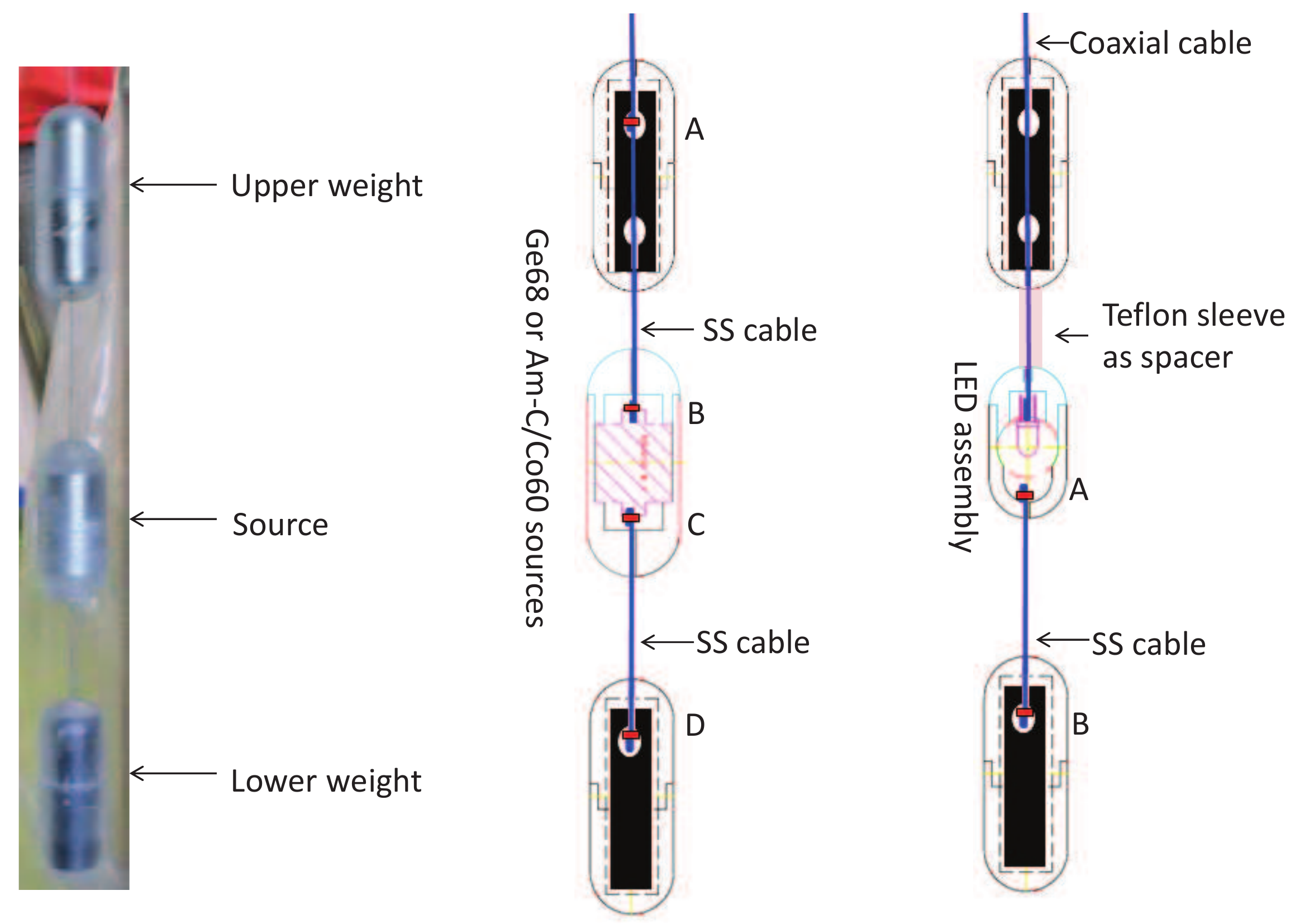}
\par\end{centering}
\caption{\label{fig:source_assembly_pic} Left: picture of a complete source
assembly. Middle and right: illustration of cable attachments for radioactive
and LED source assemblies, respectively. Red spots indicate
cable attachments. See text for details.
}
\end{figure}

Details of the cable attachment scheme is illustrated in 
Fig.~\ref{fig:source_assembly_pic}. 
For strength considerations, we 
require that the cable attachments must only be constrained by 
stainless steel materials, not by the acrylic shell\footnote{The only
exception we made is to the bottom weight for the LED source, which 
is attached to the acrylic shell to the LED. We note, however, that 
a breakage of the LED shell with LED ball and bottom weight dropping
into the detector would not introduce fatal background problem.}. 
To avoid stainless steel
crimp cutting into the cable when improperly crimped, we adopted a method 
of looping the cable onto the attachment point, making a simple 
noose knot followed by two overhand knots, then potting the knots with 
epoxy. As shown in the middle sketch in Fig.~\ref{fig:source_assembly_pic}, 
for a radioactive source, the upper weight and the source 
are connected through the deployment cable with two attachment 
points A and B. The lower weight and the source are attached through a 
separate stainless steel cable at points C and D.  The cable attaches 
to a M3 nut at points A and D so the knot/nut can be trapped into the hole on 
the weight. For the LED source, the coaxial cable runs though the upper weight 
(no attachment) and then connects to the LED inside the acrylic shell. The 
distance between the upper weight and the LED shell is maintained by 
a 2.5-cm long Teflon sleeve. The lower weight attachment to the LED is the same
as that for a radioactive source. Separations between source to both weights
are maintained at $\sim$3 cm (tip to tip), and we keep a detailed database
of as-built dimensions. 

The entire weight of a source/weight assembly is about 110 g for the LED 
assembly, and 130 g for the other two. 

\subsection{Electronics and wiring}
\label{sec:electronics}
A large number of electronic components exists in each ACU, as summarized in 
Table~\ref{tab:elec_components}. 
\begin{table}[!htbp]
\begin{tabular}{|p{1in}|c|c|p{2in}|}
\hline 
Component & \#/ACU & wires/component & details \tabularnewline
\hline \hline
Turntable motor & 1 & 4 & bi-polar motor in serial connection, A+/A-/B+/B-\\
\hline
Turntable encoder & 1 & 8 & differential readout (3 pairs) and digital 
5 V power (1 pair)\\
\hline
Deployment axes motors & 3 & 4 & bi-polar motor in serial connection, 
A+/A-/B+/B-\\
\hline
Deployment axes encoder & 3 & 8 & differential readout (3 pairs) and 5 V power\\
\hline
Turntable limit switches & 2 & 2 & single pole single throw \\
\hline
Deployment axes limit switches & 3 & 2 & single pole single throw \\
\hline
Deployment axes load cells & 3 & 8 & isolated 5 V supply to load cell (1 pair), 
 +12/-12 V input to amplifier (1 pair), 2 pairs of redundant voltage outputs 
\\\hline
CCD camera & 1 & 4 & 12 V input power (1 pair), output (coaxial)\\\hline
LED driver & 1 & 4 & TTL input (coaxial), V$_{\rm DC}$ (1 pair)\\\hline
\end{tabular}
\caption{\label{tab:elec_components}Electronics components and wiring 
requirements for each ACU.
}
\end{table}
We shall discuss details of these instrumentation and their 
wiring I/O in this section.

\subsubsection{ACU internal electronics and wiring}
\begin{figure}[!htbp]
\begin{centering}
\includegraphics[width=4in]{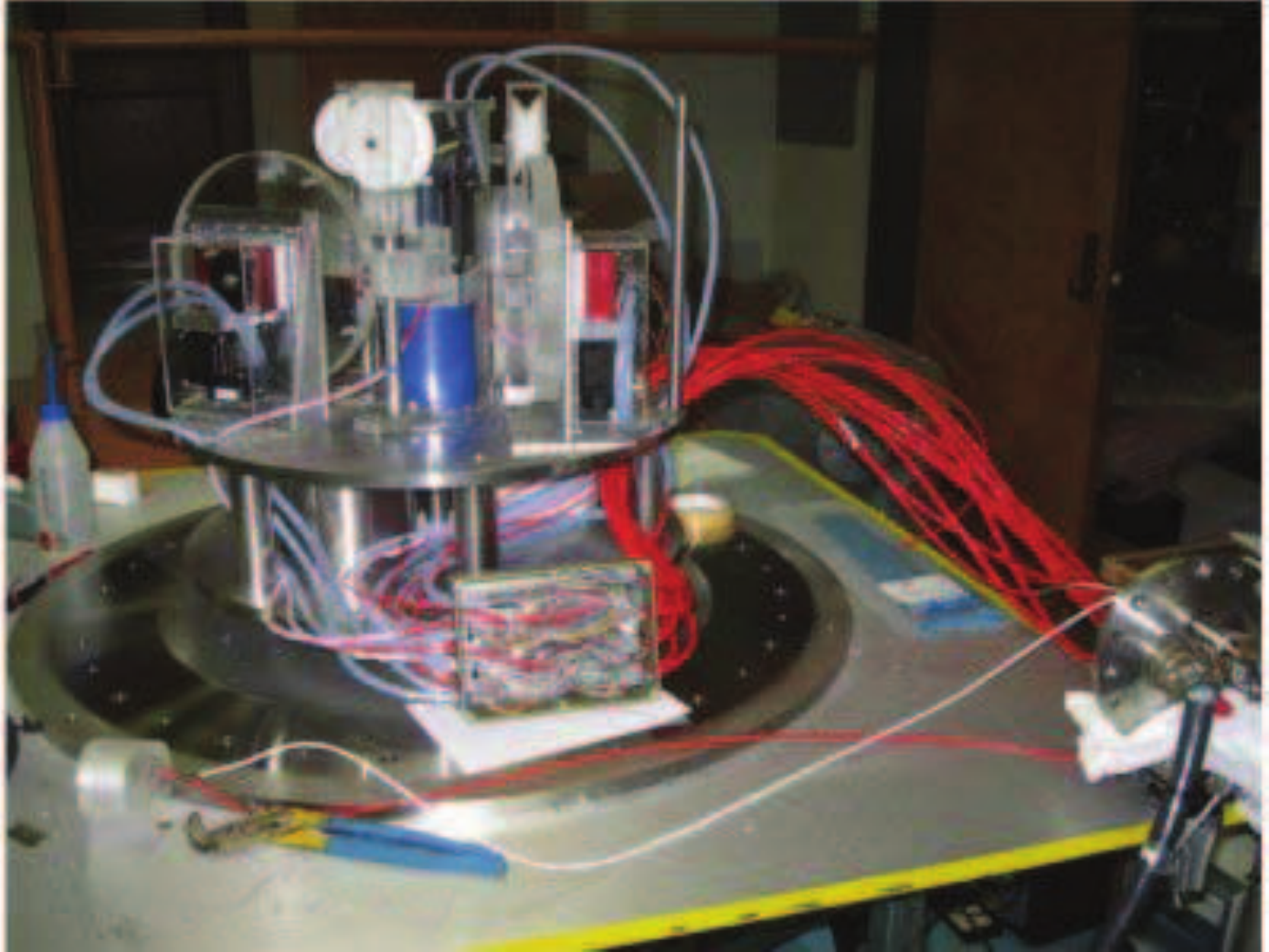}
\par\end{centering}
\caption{\label{fig:wiring_pic}ACU internal wiring picture taken during ACU
assembling. The breakout boards, 12-conductor cables (red), and the MDC
feedthrough can be seen from the pictures.}
\end{figure}

In Fig.~\ref{fig:wiring_pic}, 
a picture of wiring connections inside an ACU is shown.
A custom double-sided ISO100 vacuum feedthrough flange (MDC Inc.) hosting 
two mil-spec 52-pin feedthroughs and two isolated BNC feedthroughs is used 
on each ACU. One of the BNC connections carries the CCD camera image 
signals, and the other carries the TTL trigger signals for the LED. 
The signals from individual components get consolidated by two breakout 
boards located inside a shielded metal box mounted between the two 
turntable plates. Each board breaks into four 12-conductor (AWG 22 
shielded twisted pairs) cable, grouped by a single 55-pin female 
mil-type connector, mating to the male connector on the MDC flange. 

\begin{figure}[!htbp]
\begin{centering}
\includegraphics[width=4in]{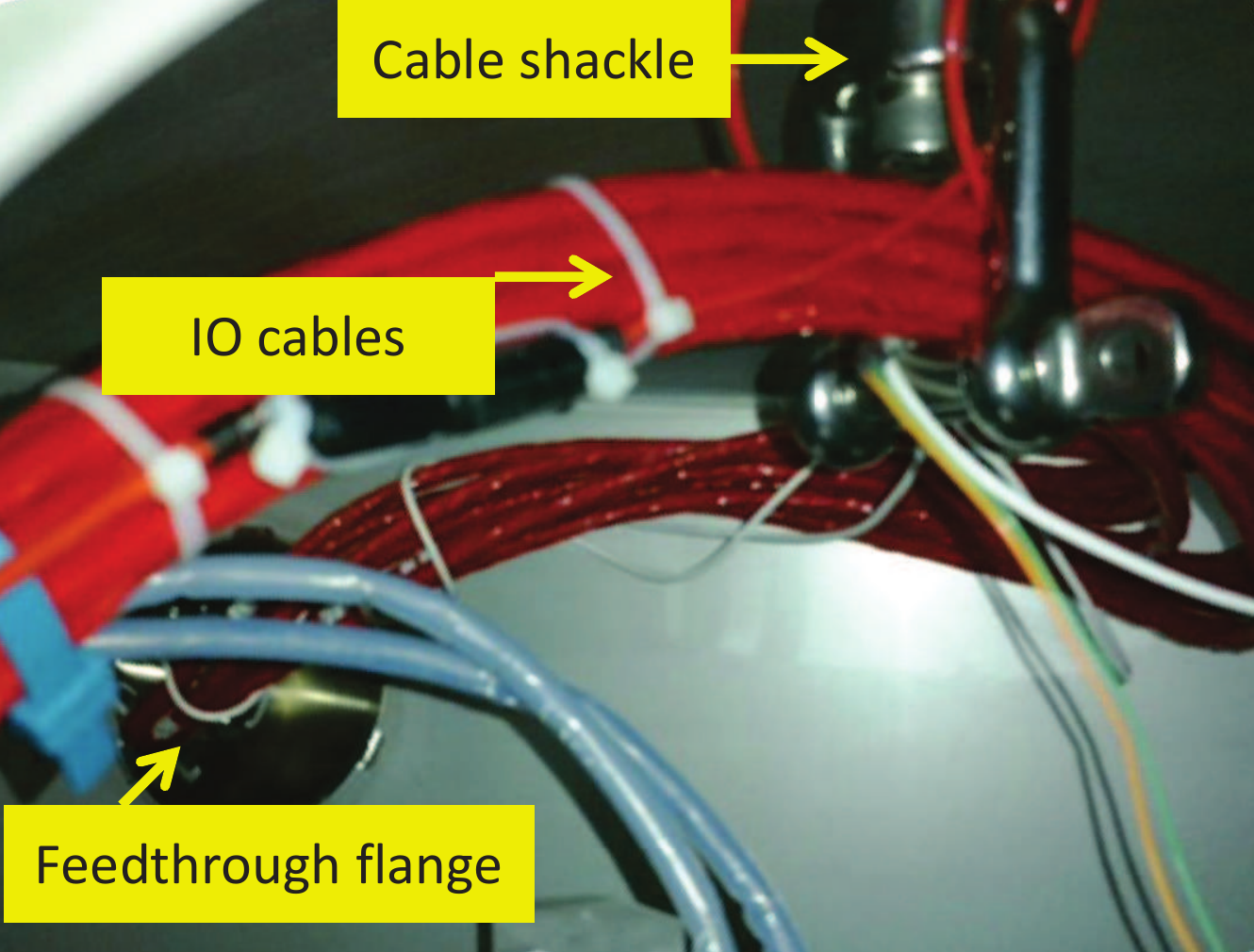}
\par\end{centering}
\caption{\label{fig:cabling_in_ACU}A picture taken when ACU internal cable
connection is made to the feedthrough flange.}
\end{figure}

A cable guiding post is mounted on the turntable.
From the breakout box, all 12-conductor cables are cable-tied to this post 
running upward. The cables 
then run through a rotatable shackle mounted on the underside of the 
bell jar dome, then go horizontally to connect to the feedthrough flange. 
This routing scheme, shown in Fig.~\ref{fig:cabling_in_ACU}, 
ensures that cables do not run into other mechanical components 
when the turntable is rotating.

\begin{figure}[!htbp]
\begin{centering}
\includegraphics[width=4in]{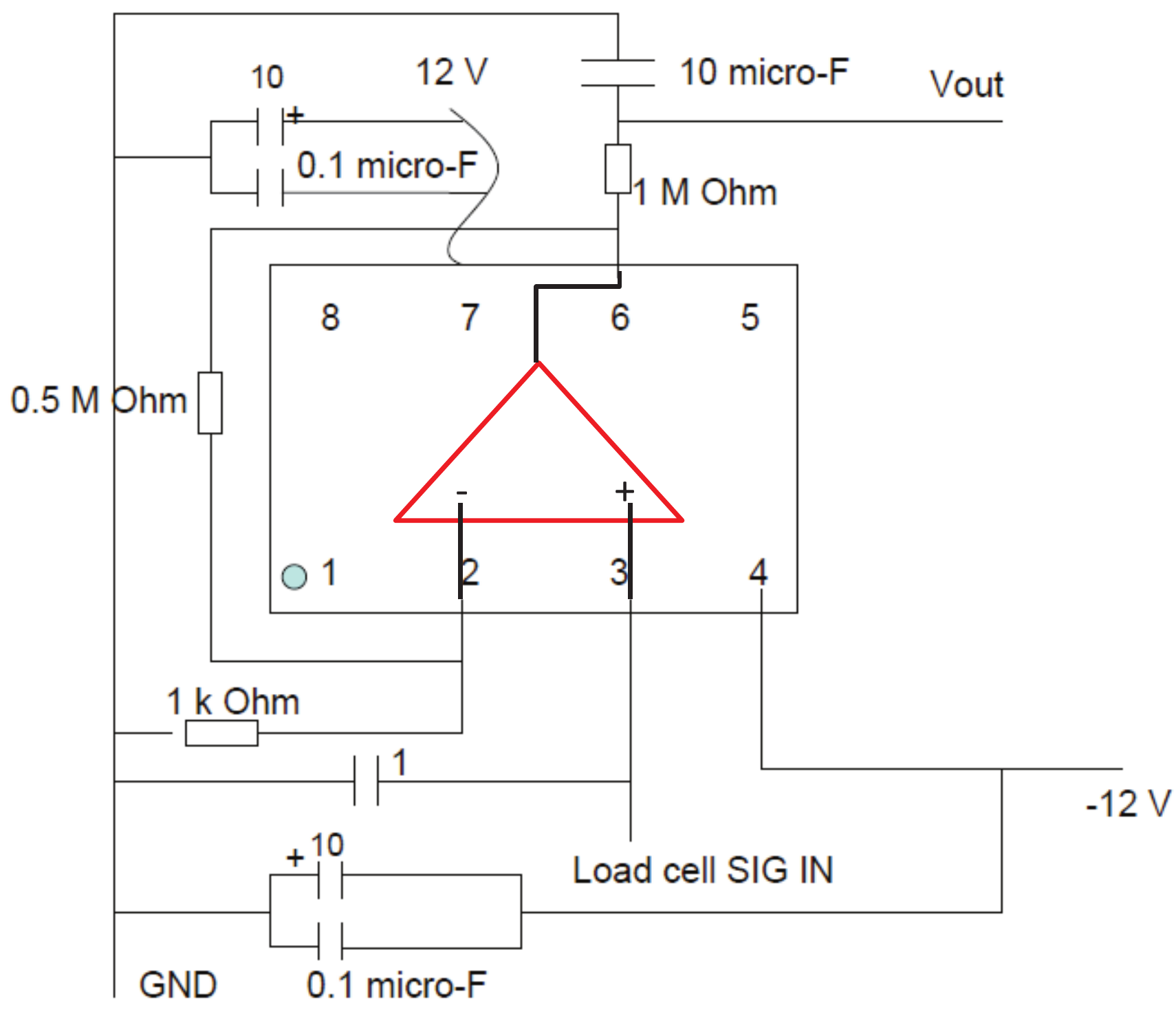}
\par\end{centering}
\caption{\label{fig:loadcell_opamp} Schematic diagram of the load cell amplifier
circuitry on the breakout board.}
\end{figure}

Most signals simply pass through the breakout boards. The load cell signals 
are exceptions, as the raw signals from them are of the order of mV 
(raw sensitivity = 2 mV/lb with 5 V power input). To avoid noise pickup 
 (which may swamp the raw signal), these signals have to 
be amplified in-situ before getting transmitted to the readout 
electronics.  Non-inverting amplifiers were 
implemented on the breakout boards using op amps (AD843) 
operating under negative feedback mode (Fig.~\ref{fig:loadcell_opamp}). 
Capacitors at the power inputs as well as at the output are essential to 
suppress noise pickup during motor movement. It was also found that the 
5V power GND to the load cell has to be isolated 
from the analog ground (defined by the $\pm$12 V GND to the opamp). 
To avoid loss of load cell output due to op amp failure, two redundant 
amplifier circuitries were implemented for each load cell.
 
\subsubsection{External electronics and wiring}
The ACUs are mounted on the lid of the AD inside the water pool during 
normal operation. ACU cables ($\sim$60 m) 
have to go all the way from the MDC feedthrough 
flange to the control electronics and computers located in the 
electronics room. Under the water, cables are sealed from the water 
by a stainless steel dry pipe system (Sec.~\ref{sec:dry_pipe_leak_check}). 
For each ACU, 
two 48-conductor cables (24 individually shielded twisted pair, AWG22) 
and two RG58 coaxial cables connect to the two mil-spec and two BNC 
feedthroughs on the MDC flange respectively.

\begin{figure}[!htbp]
\begin{centering}
\includegraphics[width=4in]{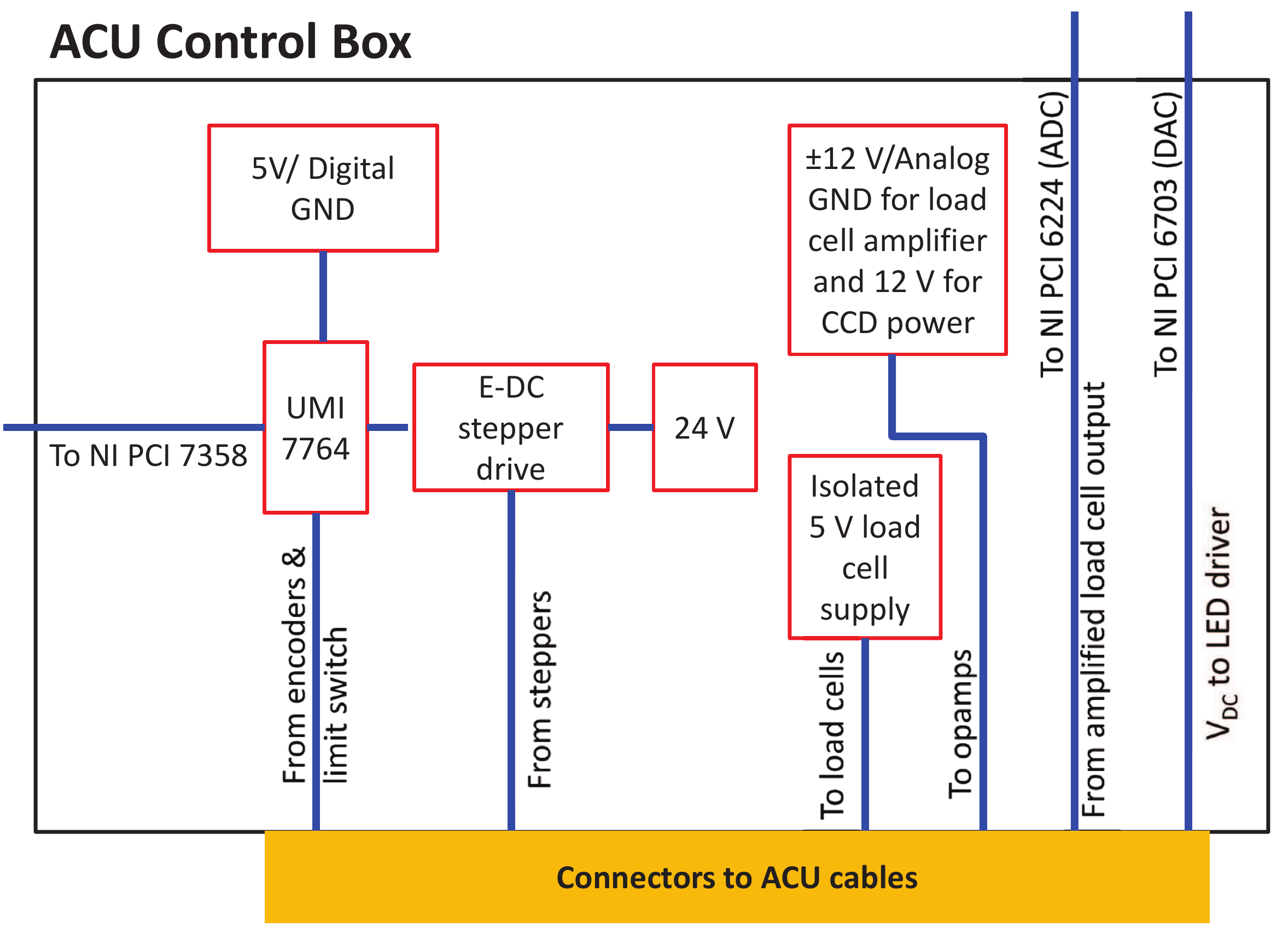}
\par\end{centering}
\caption{\label{fig:control_box} Schematic of the ACU control box.}
\end{figure}

The central hub for the ACU cables in the electronic room is a 4-U 
power/signal distribution box, one for each ACU, known as the ACU control box.
A simplified schematic diagram of the box is shown in 
Fig.~\ref{fig:control_box}. The stepper
motor wires connect to its driver (E-DC by Parker Inc.) first. The power to 
the driver is supplied by a 24 V DC power supply. As mentioned in 
Sec.~\ref{sec:failure_protection}, output current limits on these 
drivers were set by an adjustment resistor 
to limit the output torque from the stepper motors. 
An interface board 
(National Instruments UMI 7764) collects wires from the E-DC driver,
the optical encoder, as well as the limit switches, and consolidates
all motion I/O signals through its 68-pin connector, to be connected 
to the National Instrument motion control card PCI 7358. The control box also
hosts three independent isolated 5 V power supply for the three load cells\footnote{Crosstalks between load cell outputs were found when three load cells 
shared the same 5 V power supply.}, as well as the $\pm$12 V power for 
the op amp and the CCD camera. The amplified load cell signals pass through 
the control box going into the ADC (National Instruments PCI 6224), whereas 
a DAC (National Instruments PCI 6703) passes a DC voltage through 
the ACU control box to the LED driver. 

\begin{figure}[!htbp]
\begin{centering}
\includegraphics[width=3 in]{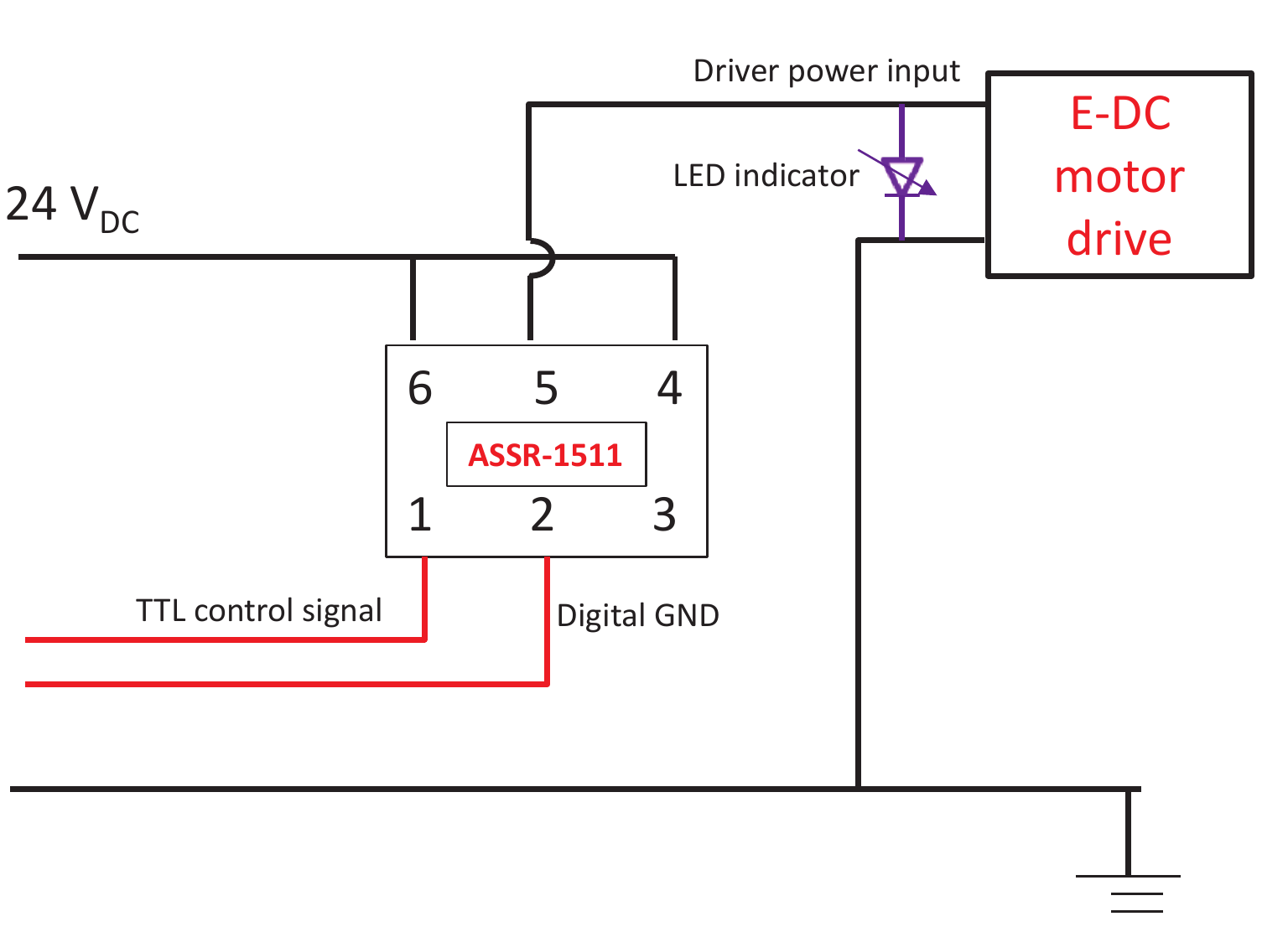}
\par\end{centering}
\caption{\label{fig:relay} Schematic of solid state relay circuitry in the 
control box to control power to the stepper motor drive.}
\end{figure}

The motion control card National Instruments PCI 7358 also contains several 
digital outputs (TTL). To avoid stepper motors introducing noises in
detectors, solid state relay chips (ASSR-1511, schematics shown in  
Fig.~\ref{fig:relay}) are implemented in the control box to 
allow remote turning off of all motor driver's power when the source is 
deployed in position\footnote{The worm gear ensures source position be 
locked during power loss.}. The same relays are used in the control box 
to switch on/off the CCD camera, LED control voltage etc. 

The TTL trigger signals for the LED are generated by 
National Instruments PCIe 6535 DAQ card,
which get carried by a RG58 coaxial cable going directly to the ACU. A copy 
of this TTL trigger gets fed into a TTL-LVPECL converter, which 
then gets input to the local trigger board (LTB) of the front end electronics.
During the LED calibration 
runs, the LTB only triggers on this external trigger. The 
signal cable from the CCD camera, also RG58, runs into a video 
IP PCI card (Q-see), video streams of which then get displayed on the 
computer.

Commercial PCI extension crates (Magma), a 7-slot one for each near site, 
and a 13-slot one for the far site, host all the PCI cards mentioned 
above. A server (Supermicro) with dual power supply 
running WindowsXP SP3 and LabVIEW 8.5  
is installed in each experimental hall,
communicating with all the automation hardware and sensors through the 
PCI crate. UPS power is in service to the computer and Magma crate to 
ensure that the system can shut down properly during power outage. 

\subsubsection{Groundings}
An effective practice to avoid ground loops in electronics is to derive 
all grounding connections from a single point. Efforts have been made to 
ensure that the electronic ground of all PMTs in the AD is 
defined {\underline{only}} by the Cu grounding bar
in the electronic room (known as the clean ground). 
The stainless steel tank is connected to the clean ground via a dedicated 
grounding braid and is insulated from everything else. 
The electronic components inside the ACUs 
follow the same guideline. Their grounds 
are defined by the control box ground (shorted to clean ground in the 
electronic room)  via individual grounding wires inside the 
48-conductor ACU cable, and electrically insulated from the bell jar and 
turntable.  The shielding braids of all twisted pairs between 
the ACU and control box are grouped together, shorted to the clean ground 
at the control box end, and cut at the ACU end. 

\subsection{Design of the control software}
\label{sub:controlware}
% I have made changes to this section -- raymond
In the design of the control software, balancing safety and automation 
has always been the main focus. From the safety point of view, the 
control software has to prevent or stop operations which can cause 
harm to the system, while, to achieve genuine automation, it 
cannot completely rely on human intervention when potentially 
dangerous situations arise. However, whenever there is a conflict, 
safety always comes first at the expense of automation. 

\subsubsection{Design}

The control software consists of several application modules written 
in LabVIEW. This modular (as opposed to monolithic) nature of the software 
minimizes coupling between parts and simplifies the customization for 
each site. The central piece is the Main Program which consists of 
two independent but coupled parts (or loops). Each performs a crucial 
function: the ``Monitor Loop" periodically monitor sensor readings 
and the ``Control Loop" direct signals to the electronics to control 
source motions. The Main Program also provides the main interface for 
user operation. The Data Fetcher fetches readings from the sensors, 
e.g. load cells, encoders, etc., and provides such readings to the 
main program. The Watchdog ensures that the Main Program always duly 
performs its monitoring function. The Distributed Information Management 
(DIM) Communication Module relays information between the main program 
and the data-acquisition system (DAQ) over TCP/IP, enabling automatic 
deployment. The detailed workings of the software will be discussed below 
from a functional point of view.

\begin{figure}
\centering \includegraphics[width=0.8\textwidth]{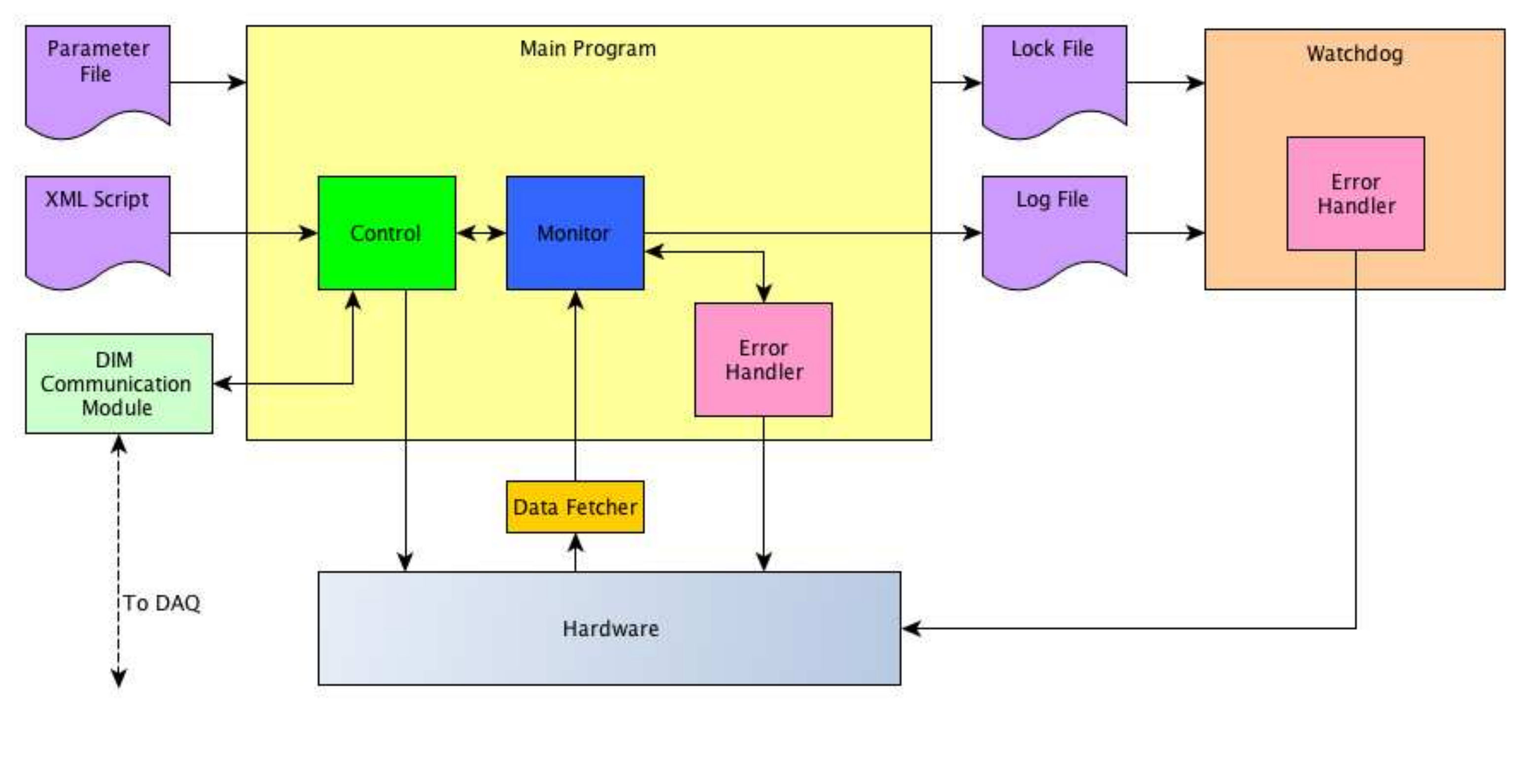}
\caption{Schematic of software structure.}

\label{fig:contro-schematic} 
\end{figure}

\begin{figure}
\centering \includegraphics[width=0.8\textwidth]{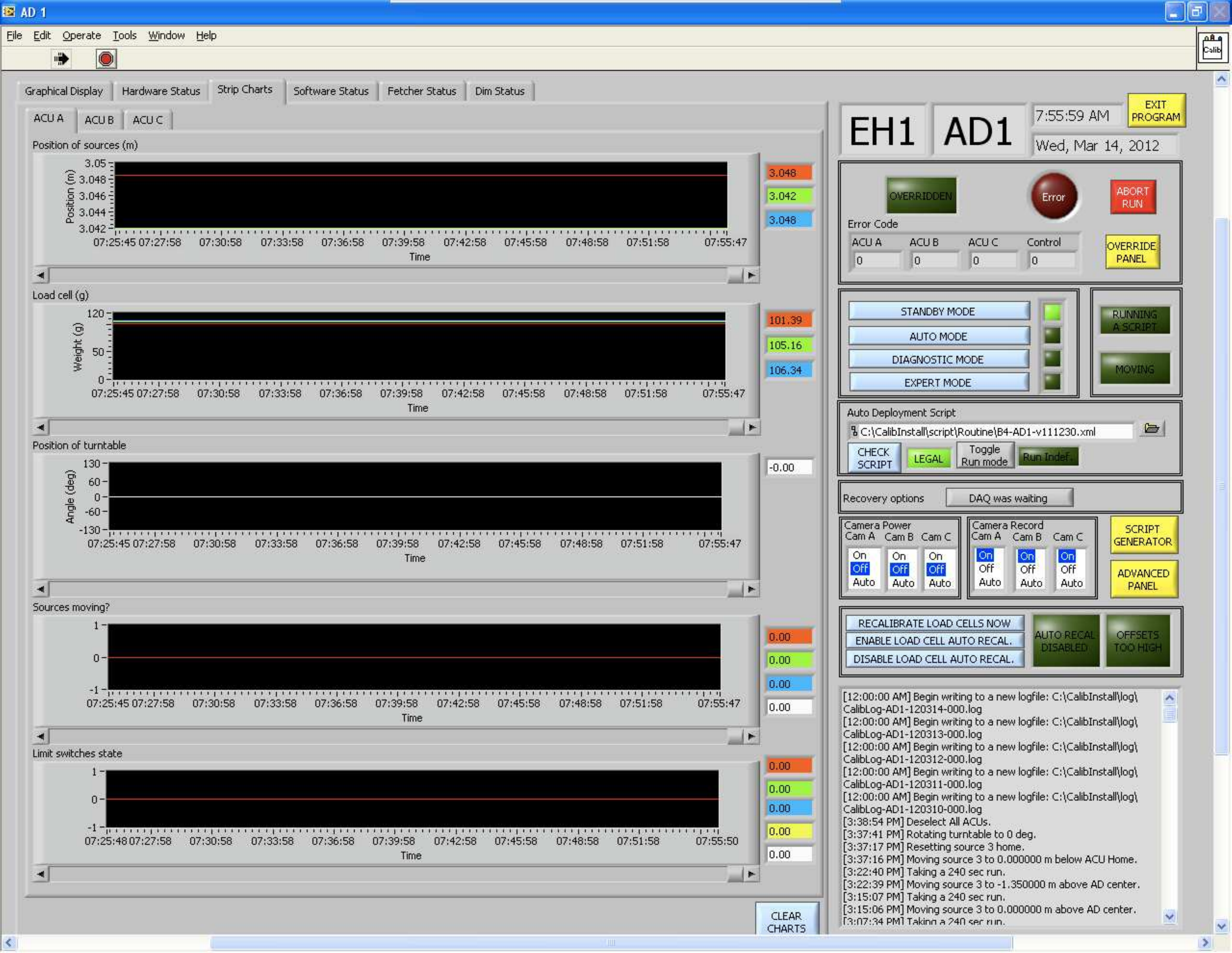}
\caption{Screenshot of the Main Program.}

\label{fig:screenshot} 
\end{figure}

\subsection{Monitoring}
Involved in the monitoring function of the software are the Monitor 
Loop in the Main Program, the Data Fetcher and the Watchdog. The 
Data Fetcher fetches readings from all sensors 
(See Table \ref{table:sensorreadings} ) from the hardware 
at the highest possible frequency allowed under computer and 
hardware constraints. The readings are then transmitted to 
the Main Program via a local DataSocket server. On receipt of 
the readings, the Monitor Loop will look for signs of danger and 
issue alarms if any such signs are observed 
(see Table \ref{table:acualarms}). When an alarm is issued, 
the Main Program would signal all motors to stop and would then 
power them down via the relay circuitry in Fig.~\ref{fig:relay}. 
After processing this set of readings, the 
Main Program would then signal the Data Fetcher to clear the 
latched set of readings and read in another set. Such a cycle 
typically runs at 1 to 2 Hz.

\begin{table}
\centering %
\begin{tabular}{|l|c|c|c|}
\hline 
Variable name  & Axis  & Type  & Unit \tabularnewline
\hline 
Stepper count  & All  & Int  & Counts \tabularnewline
Encoder count  & All  & Int  & Counts \tabularnewline
Load cell reading  & All  & Double  & Volts \tabularnewline
Motion status  & All  & Bool  & - \tabularnewline
Reverse limit switch status  & All  & Bool  & - \tabularnewline
Forward limit switch status  & Turntable  & Bool  & - \tabularnewline
\hline 
\end{tabular}\caption{Sensor readings from each ACU.}

\label{table:sensorreadings} 
\end{table}

\begin{table}
\centering %
\begin{tabular}{|l|p{1.5 in}|c|}
\hline 
Bit  & Name  & Remarks \tabularnewline
\hline 
1  & Turntable stepper/encoder mismatch  & Tolerance = 0.6 deg\tabularnewline
\hline
2-4  & Source 1-3 stepper/encoder mismatch  & Tolerance = 2.5 mm\tabularnewline
\hline
5-7  & Source 1-3 reaches maximum depth  & IAV bottom or OAV bottom\tabularnewline
\hline
8-10  & Source 1-3 load cell out of limit  & below 50\% of or 300 g above nominal weight\tabularnewline
\hline
11  & Source moves when turntable is misaligned  & Prevent misoperation\tabularnewline
\hline
12  & Turntable moves when some source is deployed  & Prevent misoperation\tabularnewline
\hline
13  & More than one motor move simultaneously  & Prevent misoperation \tabularnewline
\hline
14-18  & Inconsistent status  & Ensure internal consistency\tabularnewline
\hline
19-21  & Load cell saturated  & Load cell offset $<$ -9.0 V \tabularnewline
\hline 
\end{tabular}\caption{Alarms that can be issued by the Main Program}

\label{table:acualarms} 
\end{table}

To provide an additional layer of security, the Monitor Loop 
is constantly watched over by the Watchdog, making sure that the 
Main Program is running and updates the log file at a fixed frequency 
(typically 1 Hz). An alarm will be issued when this frequency is not met.

\subsubsection{Control}
\label{sec:software_control}

\begin{figure}
\centering \includegraphics[width=0.8\textwidth]{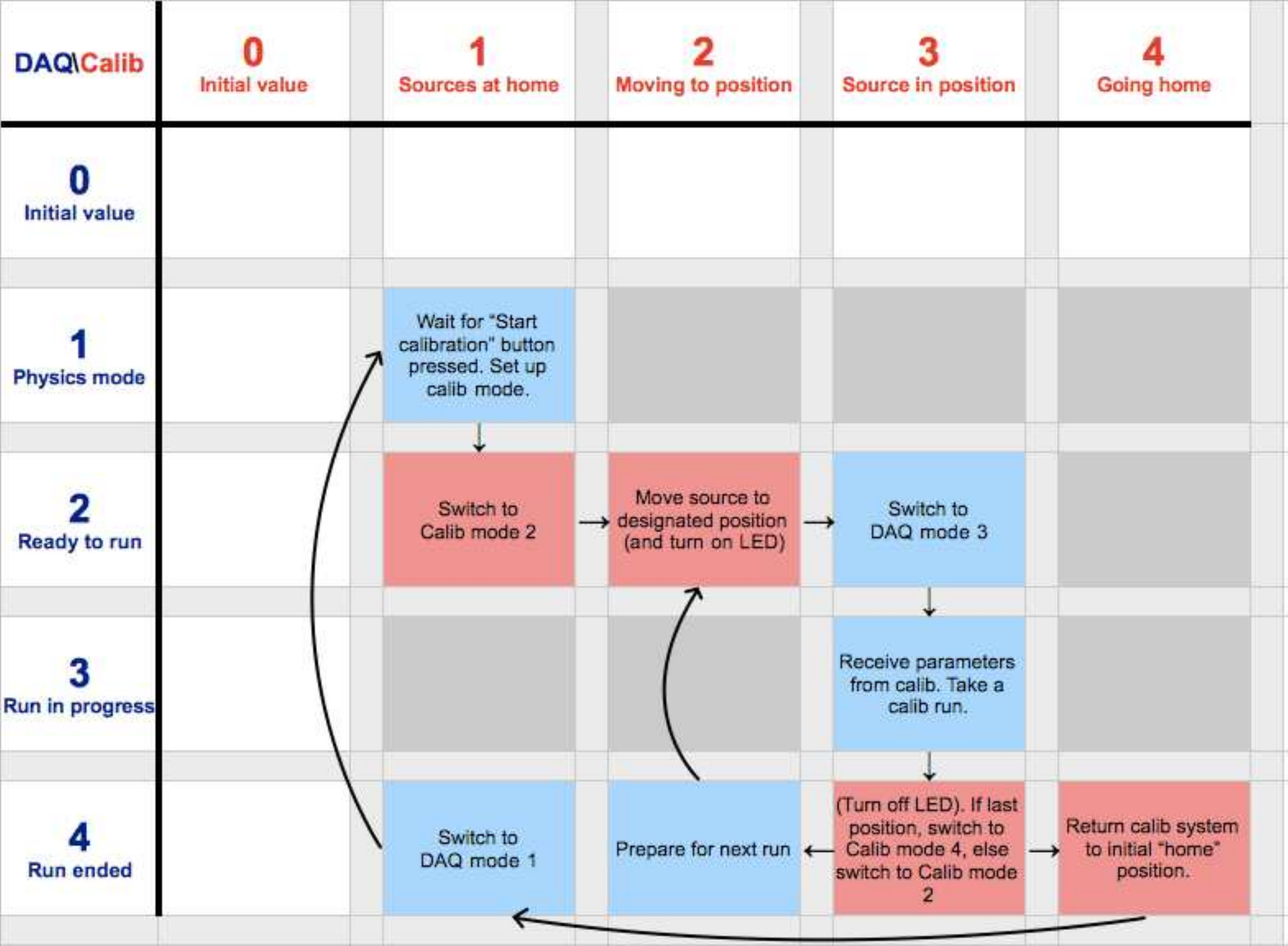} \caption{Communication protocol between the control software and DAQ. See text for details.}

\label{fig:daqcomprot} 
\end{figure}

The Control Loop of the Main Program provides 3 modes of 
operation: Manual, Diagnostic and Auto, for controlling the 
four axes (1-3 for deployment axes and 4 for turntable) 
of motion for each ACU, and the voltage and frequency of 
the LEDs. The Manual mode is restricted to expert use for 
non-standard deployments. Any possible operation can be performed 
unless forbidden by the Monitor Loop. Unlike the Manual mode which 
requires point-and-click by the user, the Diagnostic mode and 
the Auto mode accept an XML script which specifies the sequence of 
operations to be performed.
%%(See table \ref{table:xmlcmd})
In Diagnostic 
mode, the Control Loop simply performs each operation in the XML script 
sequentially until the end of the file. 
In the Auto mode, the Control Loop and the Daya Bay DAQ system communicate
via DIM. Both systems publish their status on a dedicated DIM server, and
listen to the other side with a handshaking protocol depicted in 
Fig.~\ref{fig:daqcomprot}. 
Control Loop listens to the DAQ for the shifter's "Start" signal
before deploying any source. The Control Loop signals the DAQ to start a 
run when the source reaches the designated position and the DAQ would 
reply to the Control Loop when the DAQ run is done. The Control Loop 
then executes the next command (if there is any) in the XML script, 
performs the above ``handshake" and repeats until the end of the XML 
script, at which point the Control Loop informs the DAQ of the end of 
calibration and returns to the initial state. When an error occurs 
during Auto mode, the DAQ would be notified, and after recovery the 
calibration can resume from the place where it left off without any 
intervention from the DAQ side in most cases. 
By this means, after the shifter commences the weekly calibration program, 
the entire program (typically 5 hours by deploying three 
sources in each ACU, one at a time, 3-5 stop per round trip)
in all three halls and eight ADs are executed {\underline{simultaneously}} 
with fully automated data taking at one data run per source stop in each 
hall.

\subsubsection{Notification and Logging}

There are several channels employed in the software for notification.
When an alarm is issued, the Main Program would signal the detector-control
system (DCS) via DIM, which notifies the shifter. On the other hand,
the Main Program would also send an email notification to the experts
with a summary of alarms issued (Table~\ref{table:acualarms}) 
and recently executed commands, hence
expediting the recovery process. While all sensor readings are saved
onto the local disk, only a subset of monitoring-related readings
would be saved in the DCS database, and another subset of control-related
readings are saved to the online database via DAQ.

\section{Quality controls and position calibration}

\label{sec:mechanicalposition}
% I have made changes to this section -- raymond

\subsection{Mechanical reliability tests}
\label{sec:qa_tests}
Daya Bay experiment is planned to run for at least three years, and 
the automated calibration units would run on a weekly basis over this 
period. This amounts to at least 156 full calibration cycles for each ACU. 
We constructed in total 25 (24 and a spare) ACUs.  
To ensure the robustness of each ACU, longevity tests were undertaken at 
Caltech before shipping to Daya Bay. The longevity test involves running 
the ACU for 200 consecutive full deployment cycles. Each cycle involves 
deploying each of the three sources to a distance corresponding to the 
detector center, and then retracting to the origin. Each cycle takes about 
half an hour, and a complete longevity test for each ACU takes about four 
days. No noticeable damage has been found in any of the parts after the 
longevity test.

Two different stress tests, which emulate situations where parts of ACUs 
are damaged, were also performed. One stress test, dubbed the ``extreme test", 
targeting the limit switch and the load cell, involves forcefully pulling the 
source assembly against a disabled limit switch until the stepper counts gets 
out-of-sync with the encoder counts. This forceful pull was repeated at least 
200 times for each of the axes. The objective of this test was to make sure 
that the functionality of the limit switch and the load cells would not be 
affected under such an extreme condition. The other, called ``load test", 
aimed at ensuring secure attachment of the sources, involves hanging a weight 
of about 1 kg (max possible pull from the motor) from the source assembly for 
a duration of at least 15 minutes. No source ever failed any of these tests.

\subsection{Position calibration}
\label{sec:pos_calib}

Source deployment positions are required to have an accuracy of 0.5 cm. The
elongation of the stainless steel wire due to the weight of the sources 
is negligible, so the positional accuracy is primarily determined by the 
accuracy in the diameter of the acrylic wheel. There are also some small 
effects coming from the depth of the grooves into which the wires are wound 
and the alignment between the deployment wheel and the auxiliary wheel. 
Given that a deployment to the center of the detector typically requires 
4 turns of the wheel, uncertainty in the deployment length can be an order 
of magnitude greater than that of the wheel diameter. Therefore, a sub-mm 
uncertainty in the wheel diameter can possibly jeopardize our required 
accuracy. We devised a method to precisely estimate the 
effective diameter of the wheel. We constructed a calibration 
ruler, with accurately measured marks, aligned along the source deployment 
axis. The positions of the source can then be compared with the marks on 
the ruler. Hence, an effective diameter of the wheel can be estimated 
accurately with a linear fit. 

The calibration ruler is made of a roughly 6 m long Teflon coated stainless 
steel wire, with a 50 g weight attached to one end. Six crimps which served 
as calibration marks were attached at various measured positions on the wire. 
The ruler was then lowered into a mock-up detector, a 5.5 m long acrylic tube, 
on which we marked the positions of the crimps. We then offset the 
ruler by 200 mm, and a different set of calibration marks were 
similarly translated onto the tube. We also used the top surface of the 
turntable as a calibration point. 
%%These 13 points are listed in the first 
%%column of Table \ref{table:calpoints}.
%%Note that only the relative positions among the marks are important.
%%\begin{table}
%%\centering
%%  \begin{tabular}{| c | c | c |}
%%\hline
%%Position [mm] & Remarks \\ \hline
%%165 & Inside ACU \\
%%2000 & Set A \\
%%2200 & Set B \\
%%2299 & Set A \\
%%2499 & Set B \\
%%2798 & Set A \\
%%2998 & Set B \\
%%3197 & Set A \\
%%3397 & Set B \\
%%4496 & Set A \\
%%4696 & Set B \\
%%4996 & Set A \\
%%5196 & Set B \\
%%\hline
%%\end{tabular}
%% \caption{List of calibration marks. Positions are measured from an arbitrary 
%%reference point.}
%% \label{table:calpoints}
%%\end{table}

For each axis, the source was first deployed close to each
calibration mark. The source was then made to inch along the axis
in steps of about 0.1 mm. The encoder counts of the source motor was
recorded when the source was flush with the calibration mark. 
%%Table
%%\ref{table:typposcal} shows a typical set of encoder counts collected
%%for an ACU. 
%%\begin{table}
%%\centering %
%%\begin{tabular}{|c|c|c|c|c|}
%%\hline 
%%Calibration  & Axis 1  & Axis 2  & Axis 3  & Type of \tabularnewline
%%mark {[}mm{]}  & Encoder counts  & Encoder counts  & Encoder counts  & calibration mark\tabularnewline
%%\hline 
%%165  & 55417  & 55266  & 55599  & Inside ACU \tabularnewline
%%2000  & 670012  & 671036  & 678232  & Set A \tabularnewline
%%2200  & 736661  & 738356  & 745724  & Set B \tabularnewline
%%2299  & 770491  & 771515  & 779884  & Set A \tabularnewline
%%2499  & 837641  & 837996  & 847369  & Set B \tabularnewline
%%2798  & 937283  & 939481  & 949361  & Set A \tabularnewline
%%2998  & 1004438  & 1006133  & 1016847  & Set B \tabularnewline
%%3197  & 1072096  & 1072781  & 1085344  & Set A \tabularnewline
%%3397  & 1138242  & 1139766  & 1152996  & Set B \tabularnewline
%%4496  & 1507502  & 1509696  & 1526108  & Set A \tabularnewline
%%4696  & 1574487  & 1576180  & 1594100  & Set B \tabularnewline
%%4996  & 1673961  & 1677497  & 1695246  & Set A \tabularnewline
%%5196  & 1741117  & 1743646  & 1763237  & Set B \tabularnewline
%%\hline 
%%\end{tabular}\caption{Raw encoder counts from a typical position calibration (ACU 5A). }

%%\label{table:typposcal} 
%%\end{table}

\begin{figure}[!htbp]
\begin{centering}
\includegraphics[width=3in]{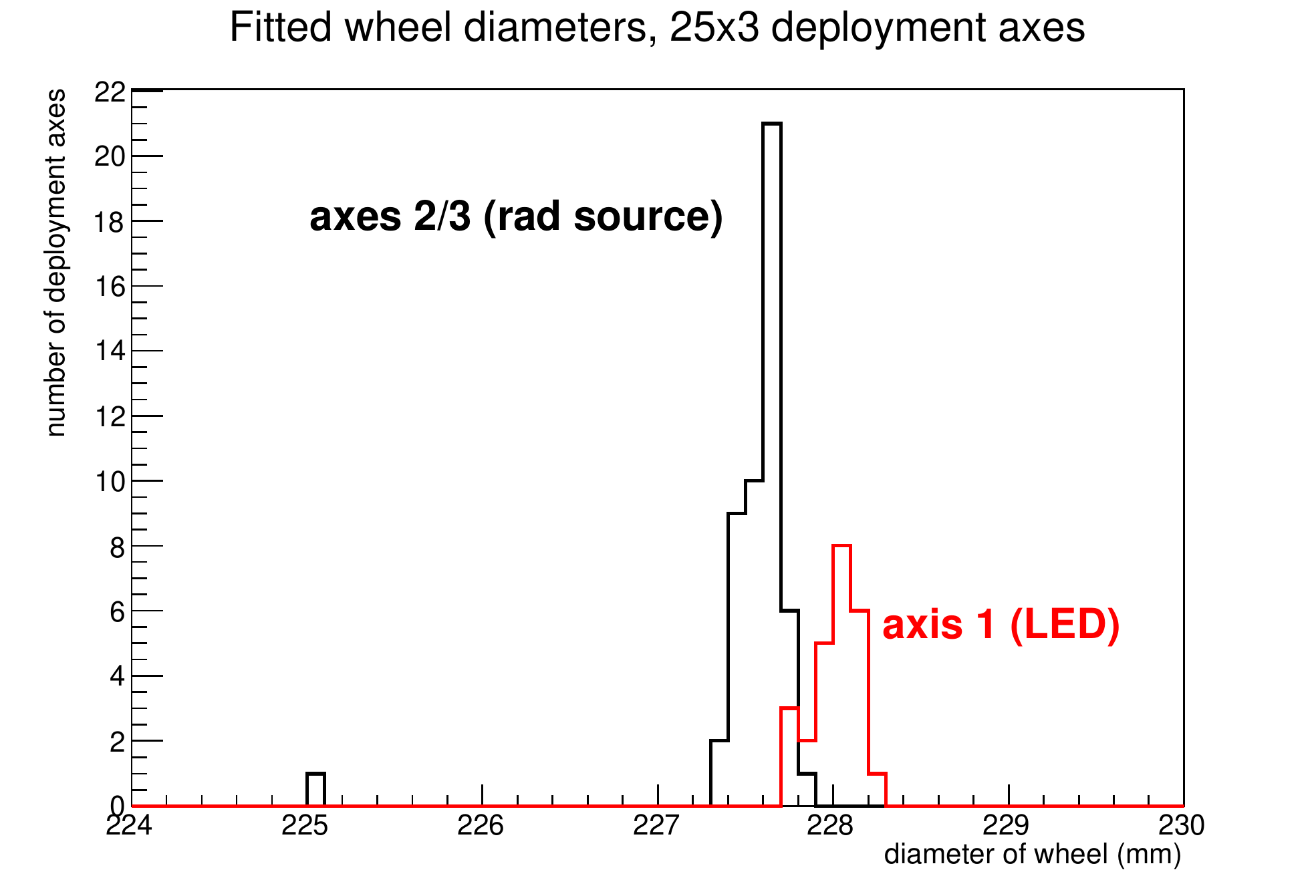}
\par\end{centering}
\caption{\label{fig:wheel_diameter}Effective wheel diameters of all 25$\times$3
ACU axes obtained by the position calibration. }
\end{figure}

The data were then subject to a linear fit of the form $y=kDx+C$,
where $k=\frac{\pi}{4000\times60}$ (4000 is the encoder count/resolution, 
and 60 is the gear ratio), x and y are respectively the
encoder counts and position of calibration marks, and the fit parameters
D and C are respectively the wheel diameter and the offset.
Fig.~\ref{fig:wheel_diameter} shows the distribution of 
wheel diameters with an average at 227.7 mm.
Most of the wheel diameters are in good agreement within 1 mm, except 
that for ACU5A axis 3, which has a $\sim$3 mm smaller diameter compared 
to the rest due to machining. One also observes
from Fig.~\ref{fig:wheel_diameter} that on average the effective 
diameter for axis 1 is $\sim$0.4 mm larger than those for axes 2 and 3, which
is consistent with the different diameters of the deployment cables 
(0.039-in for LED, 0.026-in for radioactive sources). 

\begin{figure}[!htbp]
\begin{minipage}[b]{0.45\linewidth}
\centering
\includegraphics[width=\textwidth]{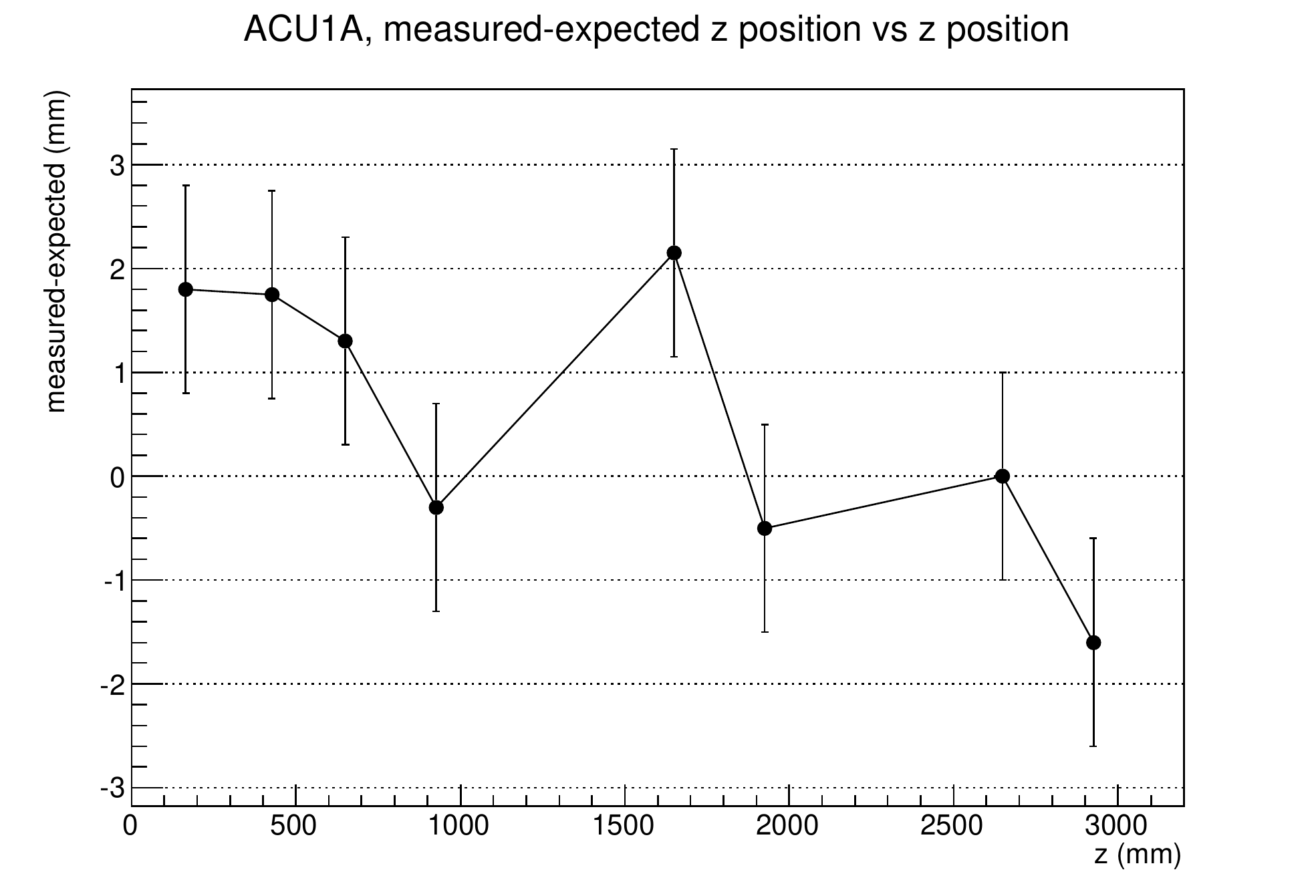}
\caption{Difference between the expected (software) and true position, ACU1A 
source 1.
}
\label{fig:pos_residual}
\end{minipage}
\hspace{0.5cm}
\begin{minipage}[b]{0.45\linewidth}
\centering
\includegraphics[width=\textwidth]{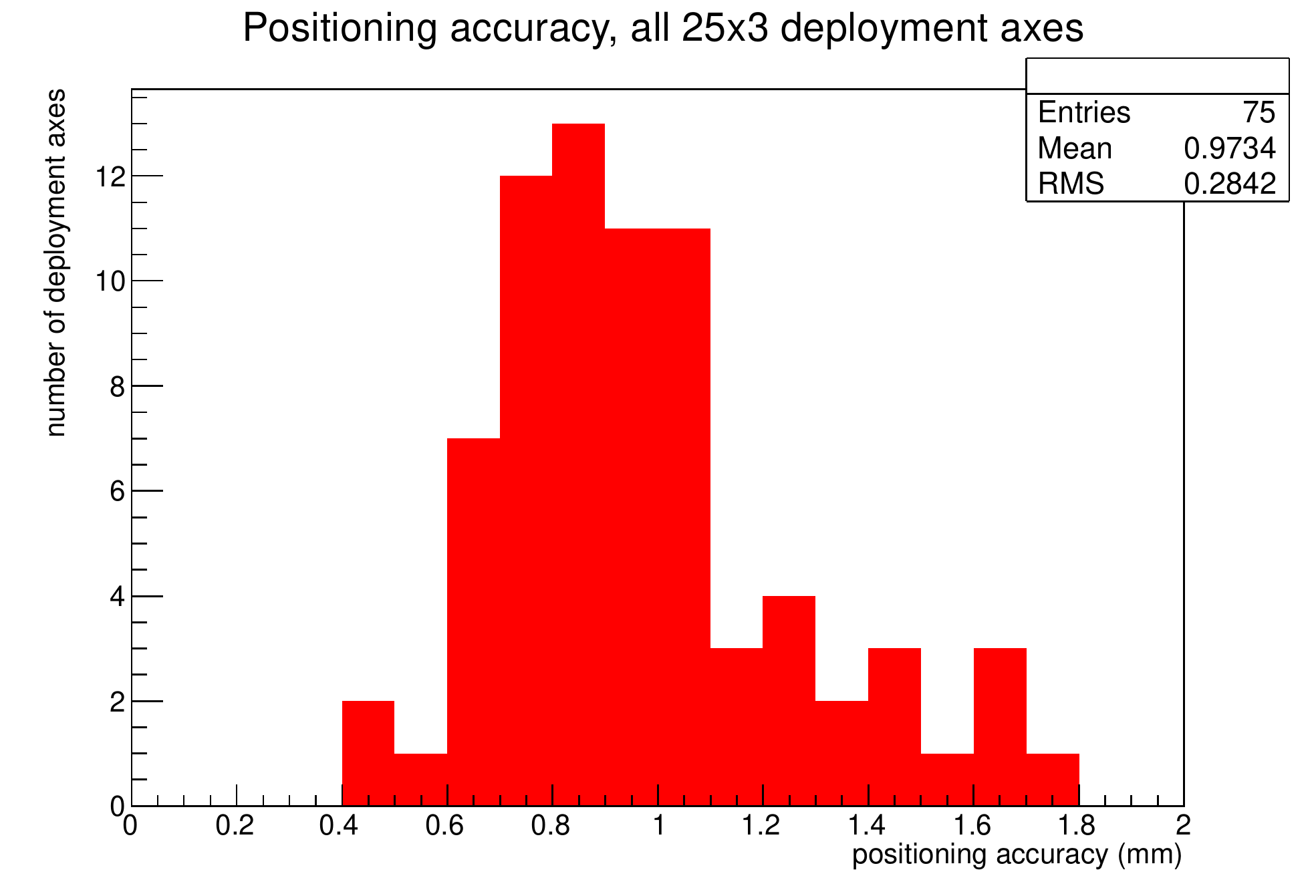}
\caption{Extracted position accuracy (based on the non-zero scattering 
in Fig.~\ref{fig:pos_residual} for all deployment axes.}
\label{fig:pos_accuracy}
\end{minipage}
\end{figure}

The parameters thus determined are used to compute the {\underline{expected}}
position at each encoder reading. For demonstration, 
differences between the expected and 
actual position on the ruler for ACU1A, deployment axis 1, are shown 
in Fig.~\ref{fig:pos_residual} as a function of vertical deployment 
length, in which the remaining scattering of 
the difference (RMS) is used as a measure of the position accuracy for 
each deployment axis. The distribution of the position accuracy
thus determined for all 25$\times$3 deployment axes (including those on the 
spare ACU) is shown in Fig.~\ref{fig:pos_accuracy}. Conservatively, we 
take $\sim2$mm as an estimate of the position accuracy of the calibration 
source to its limit switch. As a final cross check, 
obtained wheel diameters were input into the control software, and 
each source was deployed to the nominal AD center in the mock
detector. The position of the source and the marker on the ruler agree 
within 3 mm.

The absolute source accuracy of a deployed source within an AD is estimated
as follows. The position calibration demonstrated in 
Fig.~\ref{fig:pos_accuracy} 
is for the source-limit switch distance in the 
vertical direction only. Inside an AD, we must 
include the accuracy of 1) ACU mounting position relative to its support 
flange, 2) the AD verticality, and 3) the position of the 
ACU support flange relative to the general AD coordinate. The accuracy 
in 1) is $\sim$ 2mm in horizontal plane, given the size of bolt 
holes on the ACU bottom plate. The accuracy in 2) and 3) arises
from the AD survey data~\cite{ref:AD_survey_summary}, which translate 
to an absolute accuracy of $<$1 mm in x, y, and z coordinates. Combining 
everything in quadrature, we reach a final $\sim$4~mm position 
accuracy of a source in the AD.

%The results
%are shown in Table \ref{table:poscalresult}. 
%%\begin{table}
%%\centering %
%%\begin{tabular}{|c|c|c|c|c|c|c|c|c|c|}
%%\hline 
%% & ACU A  & ACU A  & ACU A  & ACU B  & ACU B  & ACU B  & ACU C  & ACU C  & ACU C \tabularnewline
%% & Axis 1  & Axis 2  & Axis 3  & Axis 1  & Axis 2  & Axis 3  & Axis 1  & Axis 2  & Axis 3 \tabularnewline
%%\hline 
%%AD 1  & 227.787  & 227.639  & 227.463  & 228.210  & 227.606  & 227.529  & 227.869  & 227.592  & 227.696 \tabularnewline
%%AD 2  & 227.796  & 227.741  & 227.491  & 227.931  & 227.611  & 227.417  & 227.755  & 227.547  & 227.535 \tabularnewline
%%AD 3  & 228.012  & 227.317  & 227.498  & 228.039  & 227.513  & 227.473  & 228.107  & 227.423  & 227.386 \tabularnewline
%%AD 4  & 227.882  & 227.538  & 227.628  & 228.025  & 227.433  & 227.418  & 228.089  & 227.512  & 227.446 \tabularnewline
%%AD 5  & 227.909  & 227.540  & 225.000  & 228.164  & 227.677  & 227.750  & 228.093  & 227.671  & 227.682 \tabularnewline
%%AD 6  & 228.053  & 227.817  & 227.644  & 228.138  & 227.684  & 227.557  & 227.933  & 227.646  & 227.683 \tabularnewline
%%AD 7  & 228.083  & 227.766  & 227.735  & 228.175  & 227.634  & 227.673  & 228.133  & 227.670  & 227.682 \tabularnewline
%%AD 8  & 228.012  & 227.571  & 227.645  & 228.107  & 227.757  & 227.632  & 227.943  & 227.677  & 227.745 \tabularnewline
%%Spare  & 227.984  & 227.679  & 227.633  & -  & -  & -  & -  & -  & - \tabularnewline
%%\hline 
%%\end{tabular}\caption{Effective wheel diameters in mm of all axes.}

%%\label{table:poscalresult} 
%%\end{table}

\section{Performance of the calibration system}
\label{sec:dayaanalysis}
% I have made changes to this section -- raymond
The performance of the antineutrino
detector with calibration sources 
has been thoroughly described in Ref.~\cite{DYBNIM12}.
In this section, we will show results more directly related to the performance
of the calibration system.

\subsection{Motion/sensor performance}
All ACUs have been fully functional since the beginning of Daya Bay 
data taking in August 2011. Alarms only occur at a 
rare rate of $\sim$ 2-3 per week for the entire system. 
All these alarms were identified 
as sensors picking up transient noise, in particular
during motor movement a load cell readout picked up noise once in a while
that triggered the alarm. 

\begin{figure}[!htbp]
\begin{centering}
\includegraphics[width=5in]{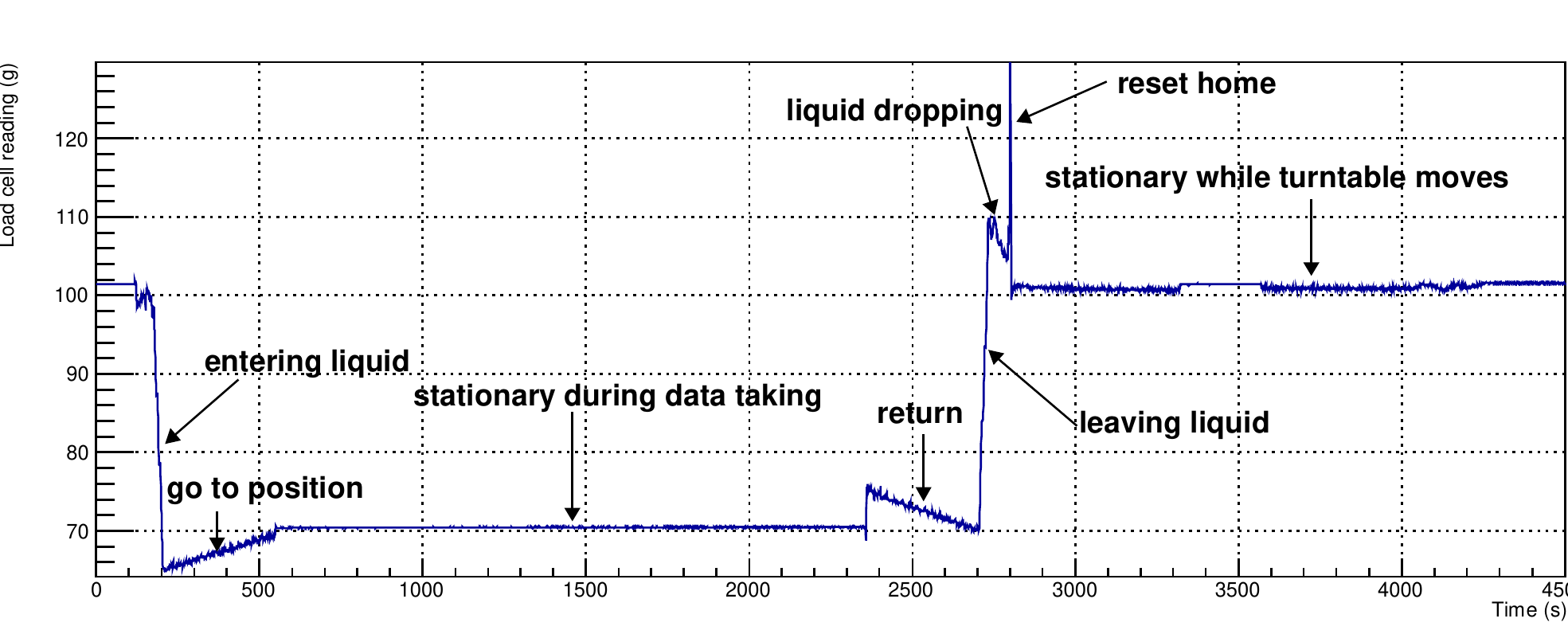}
\par\end{centering}
\caption{\label{fig:loadcell_hist}Strip chart of a load cell reading for 
a corresponding source going into and out of the liquid. }
\end{figure}

The load cell reading during a typical source deployment is depicted in 
Fig.~\ref{fig:loadcell_hist}. As the source 
goes down and crosses the liquid boundary, there is a 
change of tension in the load cell by about 30 g due to the buoyant force
applied onto the weight/source assembly. After the source is fully immersed, 
the load cell tension 
gradually increases as more cable unspools and adds to the weight felt by the 
load cell.
Once the source is at its target position, it stops and the data taking starts.
When the data have been taken, the motor reverses direction to extract 
the source from the AD. Typically there will be a sudden increase in  
load by about 10 g, 
corresponding to an additional dynamic friction that has to be 
overcome. One then sees a gradual reduction of the load due to less 
and less cable weight until the weight/source assembly crosses the 
liquid level again when the tension will recover more or less to the 
orginal level due to the loss of buoyant force. Right after the source leaves
the liquid, the liquid will be dropping back into the detector, causing
another small decrease of the load. The program will then reset the motor 
home by moving the source up till the top weight activates the limit switch, 
at which point there is a spike in the load cell. The source then moves 
down by 5 cm and get ``parked''. Even when the source is parked, 
the load cell tends to pick up some noise when other motors are moving, 
with a typical noise level of a couple of grams, as shown in 
Fig.~\ref{fig:loadcell_hist}.

\subsection{Calibration runs during the AD dry run}
\label{sec:calib_dry_run}
The LED source and the $^{137}$Cs scintillator ball played an important 
role in commissioning the detector, in particular the ``dry run'' 
before the liquid scintillator filling. 

\begin{figure}[!htbp]
\begin{centering}
\includegraphics[width=4in]{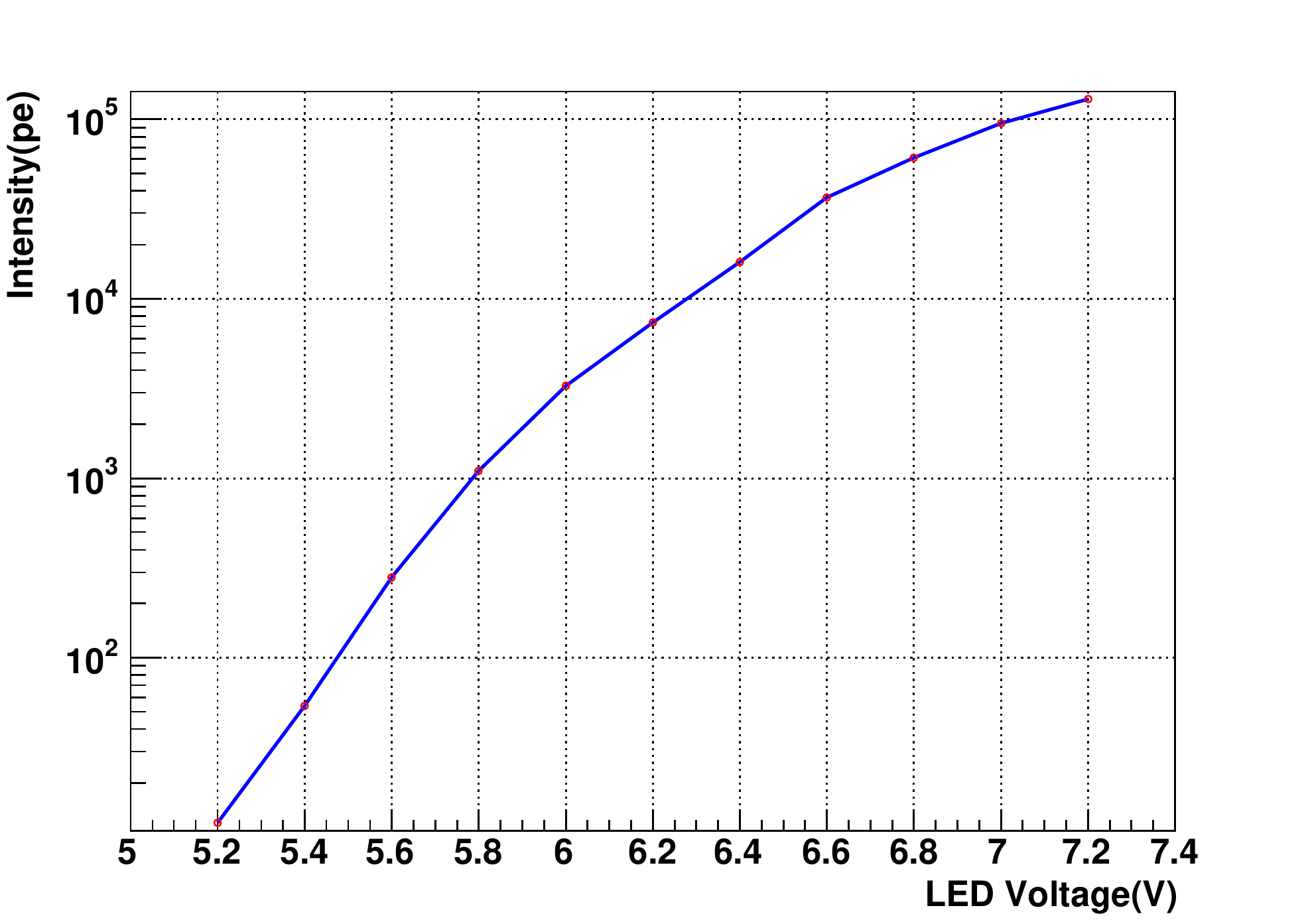}
\par\end{centering}
\caption{\label{fig:LED_intensity_scan}
Detected light intensity (in number of PE) as a function of 
LED control voltage.  
}
\end{figure}

LEDs can be used to mimic real particle interactions to study detector 
electronics. In Fig.~\ref{fig:LED_intensity_scan}, measured total 
charge from an LED intensity scan in 
an AD is shown. One sees that one could easily adjust the control 
voltage to simulate gamma-like low energy events
to muon-like high energy events, given a typical AD energy scale 
of 160 PE/MeV~\cite{DYBNIM12}.
Low energy events allow precise determination of PMT gains, 
and the high energy events allow one to study effect of PMT/electronics after 
a large pulse, e.g. baseline overshoot, ringing and recover, as well as 
retrigger issues.

\begin{figure}[!htbp]
\begin{centering}
\includegraphics[width=3in]{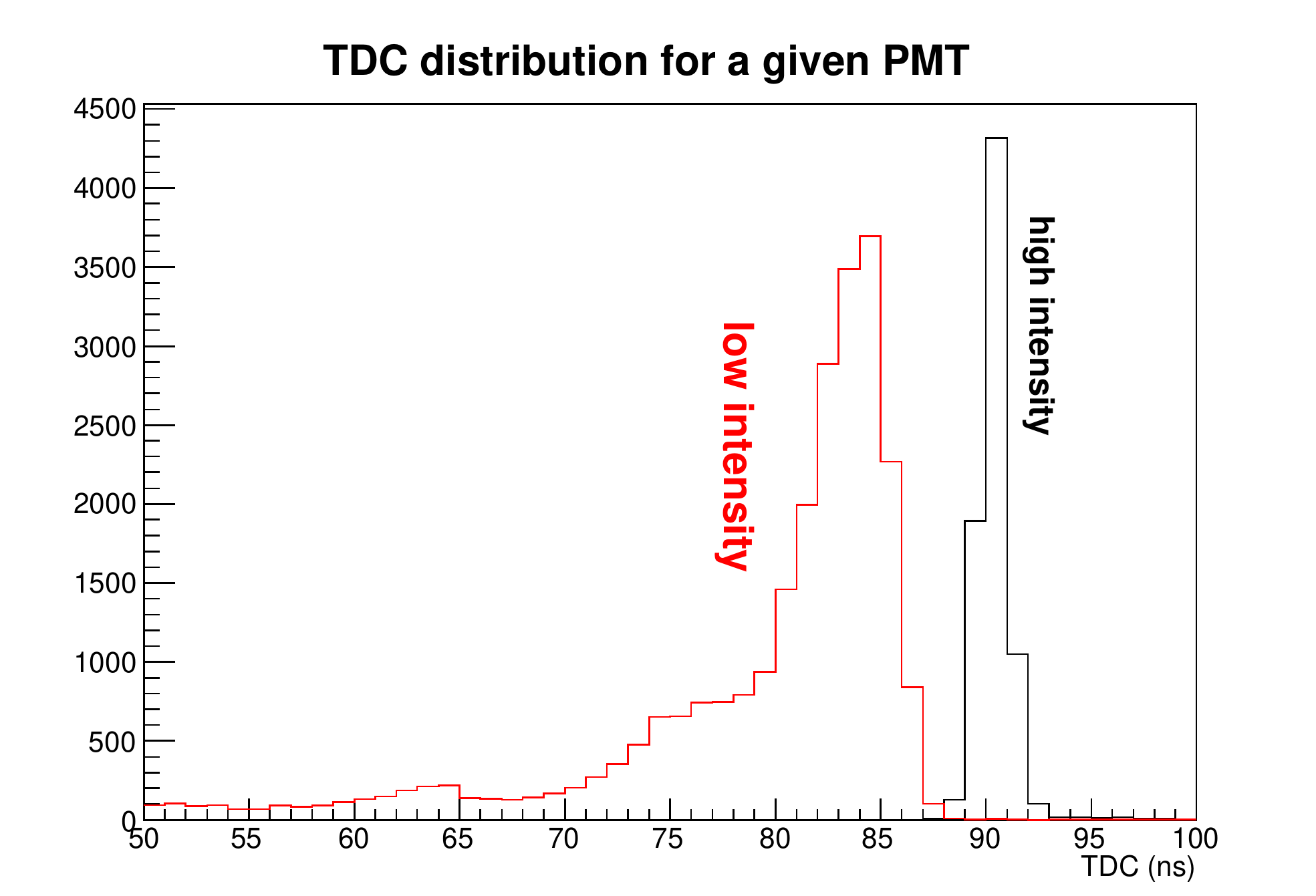}
\par\end{centering}
\caption{\label{fig:TDC_LED}TDC distribution of a PMT during a high 
(black) and low (red) LED intensity run.}
\end{figure}

%Multiple LEDs can be deployed into the detector, and be triggered in a 
%double-pulse mode to mimic the IBD prompt-delay pair with an adjustable 
%timing separation ($\Delta t$) between the flashes. The measured 
%separations between DAQ trigger time are shown in Fig.~\ref{} 
%for a series of double-pulse run with varying $\Delta t$.
Narrow timing distribution of the LED flashes relative to the trigger 
signal is an important requirement for using the LED as a timing calibrator,
in particular the rising edge of the light pulse. 
In Fig.~\ref{fig:TDC_LED}, the TDC distribution of a PMT channel (relative to the 
trigger signal generated by the TTL command pulse) is plotted for a 
high (black) and low (red) light intensity run. The sharp (RMS 0.9 ns) 
distribution in the former demonstrates that the emitted light pulse has a 
very sharp rising edge. The timing calibration data for all ADs 
were collected with LEDs under high intensity.

The TDC spectrum in the low intensity run reveals the 
overall emitted photon timing distribution. The primary pulse has a 
FWHM of 5 ns. A small late light tail of duration 
about 25 ns from the initial edge can also be observed from the 
distribution. The timing difference between the two distributions is caused 
by timing walk from the discriminator in the electronics. 

\begin{figure}[!htbp]
\begin{centering}
\includegraphics[width=3in]{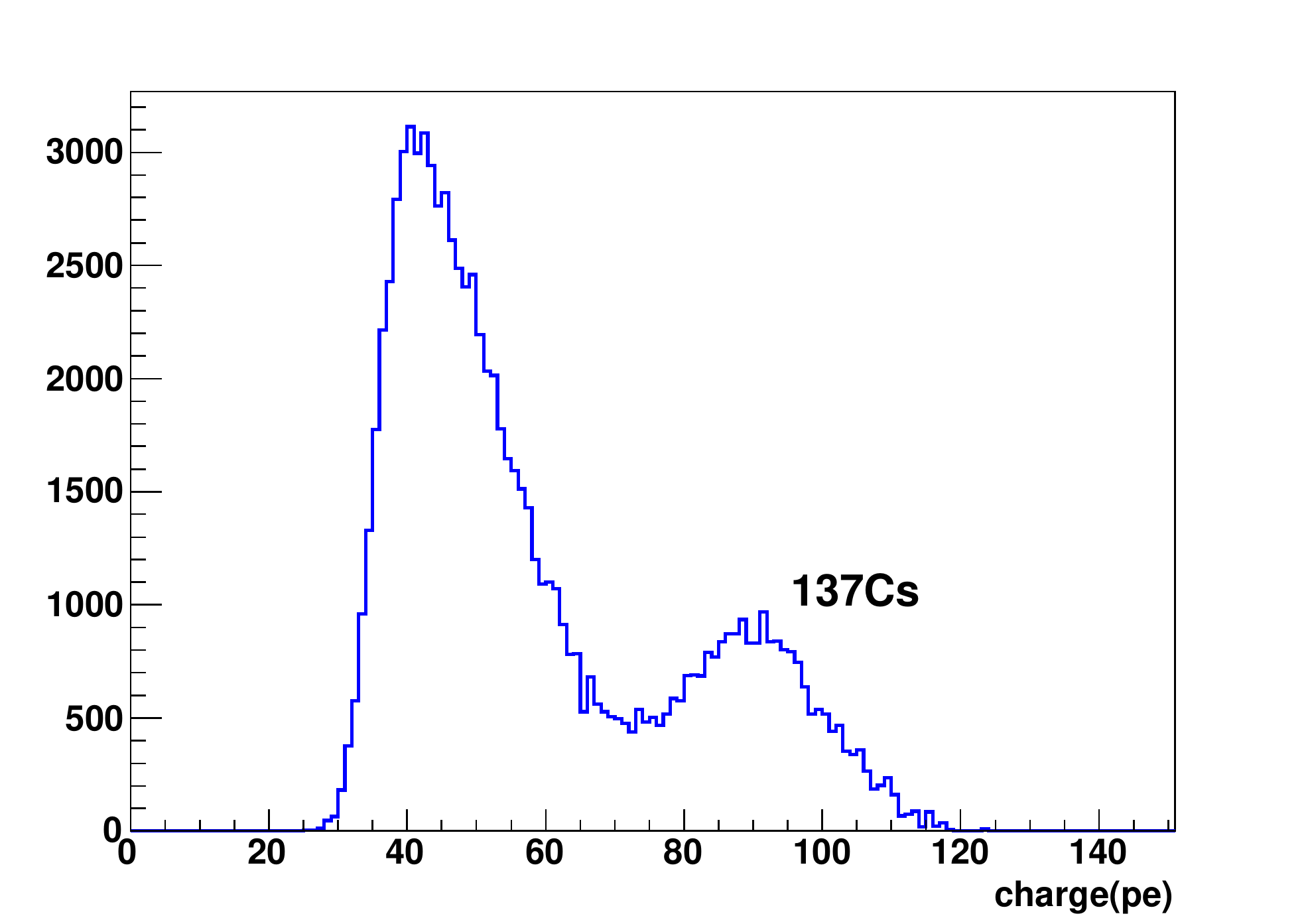}
\par\end{centering}
\caption{\label{fig:Cs_ball}$^{137}$Cs scintillator ball spectrum in a dry 
AD.}
\end{figure}

Despite these good characteristics, we discovered that 
due to different amount of coaxial 
cable winding on the spool, the LED intensity would change a few percent 
as a function of position. Therefore, to characterize the dry detector 
response, position dependent uniformity in particular, 
a $^{137}$Cs scintillator ball was designed as a stable ``candle'' 
(Sec.~\ref{sec:scint_ball}). 
In Fig.~\ref{fig:Cs_ball}, the measured 
total charge spectrum with the scintillator ball deployed at the 
detector center is shown, in which a clear conversion electron peak can 
be observed at around 90 photoelectrons. 

\subsection{Radioactive sources calibration in-situ}
\begin{figure}[!htbp]
\centering
\subfigure[AmC-$^{60}$Co] % caption for subfigure a
{
  \label{fig:spec_Co60}
  \includegraphics[width=2.5in]{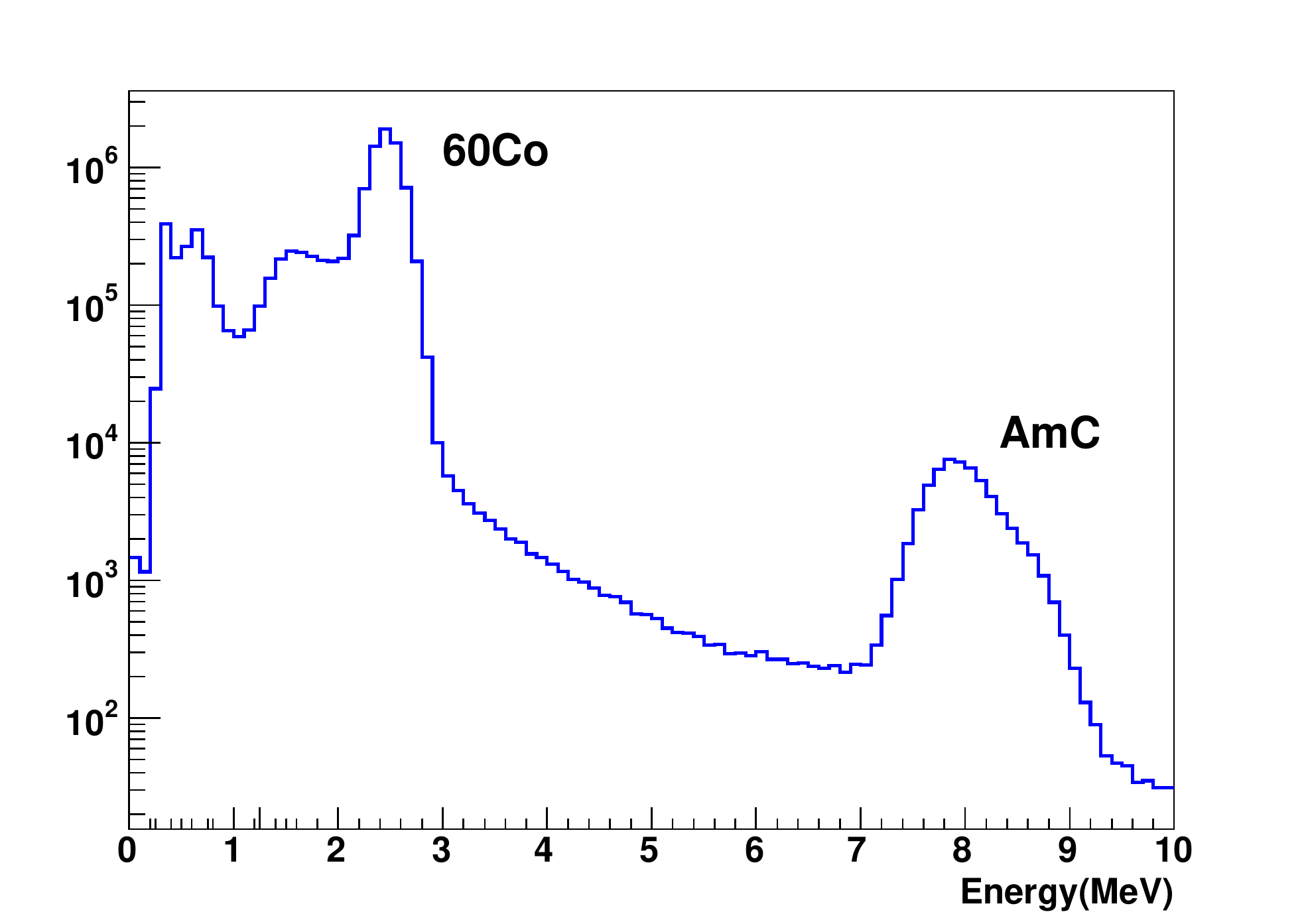}
}
\subfigure[$^{68}$Ge] % caption for subfigure b
{
  \label{fig:spec_Ge68}
  \includegraphics[width=2.5in]{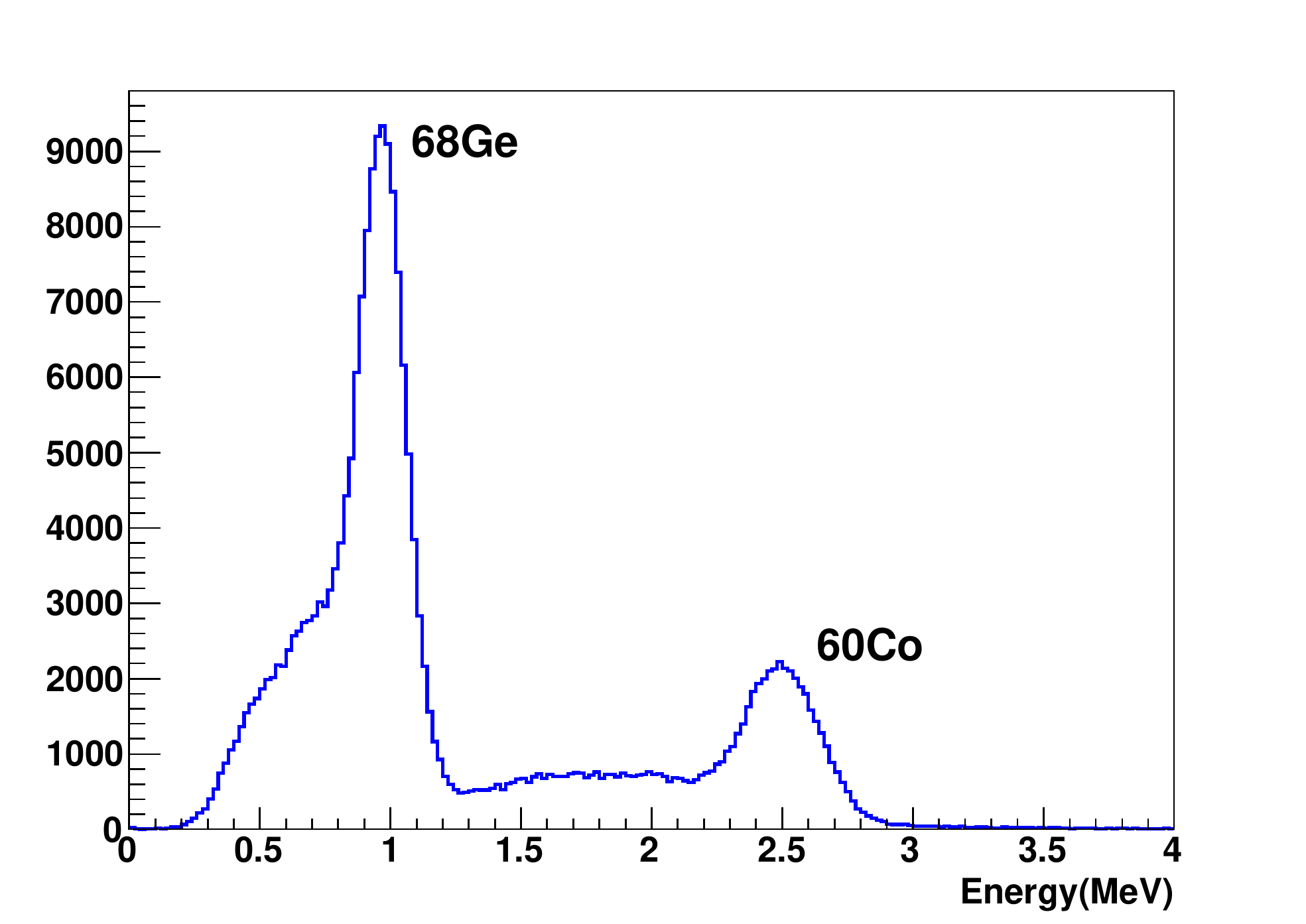}
}
\caption{       
  Energy spectrum of the $^{241}$Am-$^{13}$C/$^{60}$Co source (a) and 
  the $^{68}$Ge source (b) when deployed at the AD center.
}
\label{fig:source_spectrum} % caption for the whole figure
\end{figure}

The $^{241}$Am-$^{13}$C/$^{60}$Co and $^{68}$Ge source spectra in a full AD are shown in 
Fig.~\ref{fig:source_spectrum}. Low energy shoulders due to 
dead material on the source assembly are visible on both spectra.
The full absorption peak around 2.5 MeV in the $^{60}$Co spectrum 
is the primary method of calibrating the PE to MeV conversion factor from ADs.
The low energy peaks on the spectrum were identified 
as gamma lines from $^{241}$Am (662, 722 keV). 
The neutron-Gd capture gamma peak is also clearly observed around 
8 MeV. Since the decay rate of $^{60}$Co dominates (100 Hz 
vs 0.7 Hz of Am-C), the n-H capture from the Am-C source will produce 
negligible bias to the $^{60}$Co energy (nor can it be identified due to
the energy resolution). 
The variations of the energy scale among 
all six ADs have been controlled within 0.5\%~\cite{DYBNIM12}.
Some $^{60}$Co contaminant was found in $^{68}$Ge, which may come from 
imperfect control of the fabrication environment. However it does not 
affect effective application of this source.

%The Am-C source would emit a single neutron. It is used to study the
%Gd-neutron capture time, H to Gd ratio in Gd-LS region, as well as
%the energy scale of Gd-neutron capture peak. The Gd-neutron capture
%time of AD1 and AD2 are determined to be 28.70 $\pm$ 0.15 $\mu s$
%and 28.60 $\pm$ 0.15 $\mu s$, respectively.

%Combining all sources, the resolution of the reconstructed
%energy was determined to be $(\frac{7.5}{\sqrt{E~MeV}}+0.9)\%$. In
%addition, with gammas of various true energies from different sources,
%one can study the non-linearity (quenching in particular) of the energy
%scale. 

\begin{figure}[!htbp]
\begin{centering}
\includegraphics[width=4.5in]{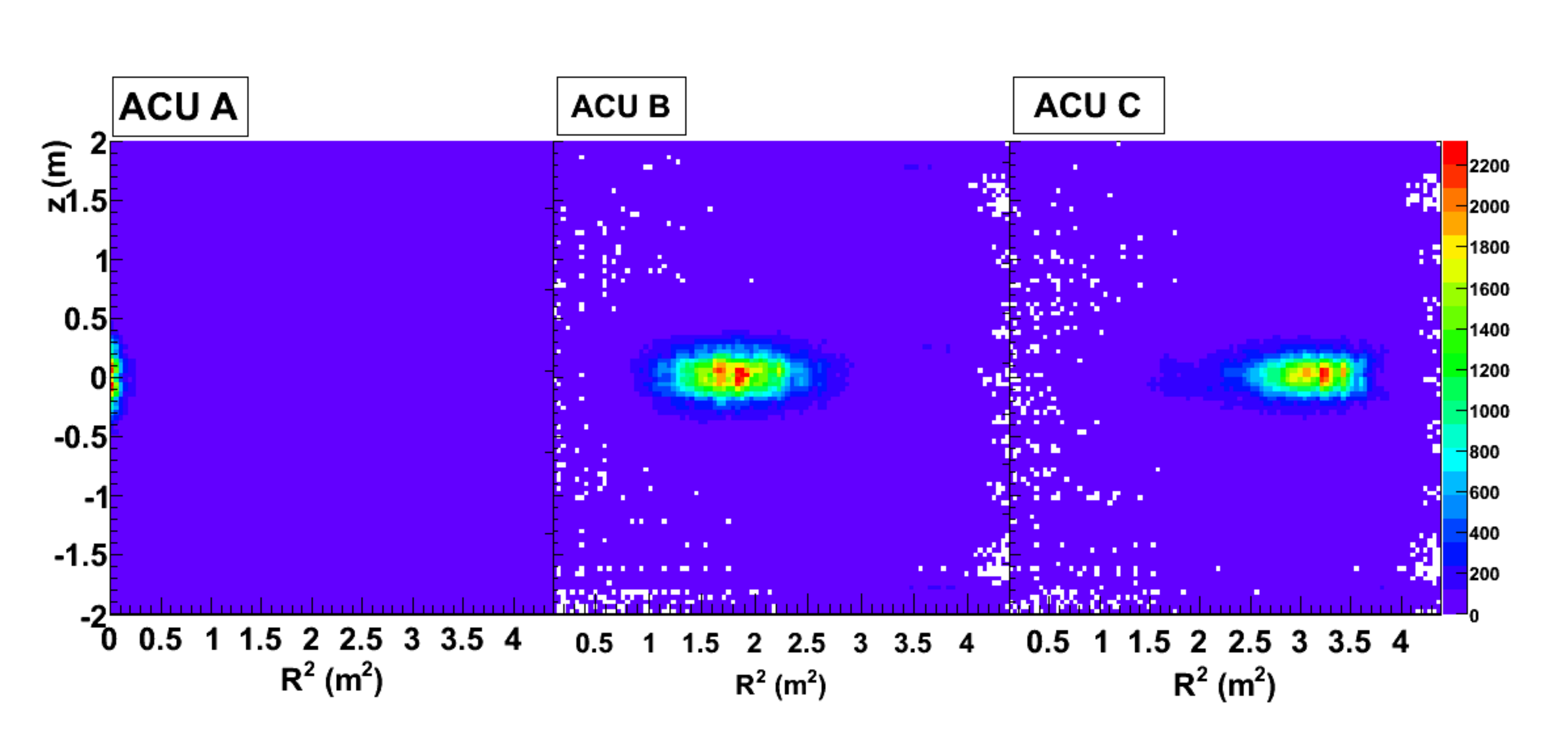}
\par\end{centering}
\caption{\label{fig:source-position}Reconstructed $^{60}$Co source position in 
a near site AD. Y axis:
z position of the source with AD center being z=0. X axis: r$^{2}$
position of the source with AD center being r$^{2}=0.$}
\end{figure}

\begin{figure}[!htbp]
\begin{centering}
\includegraphics[width=4.5in]{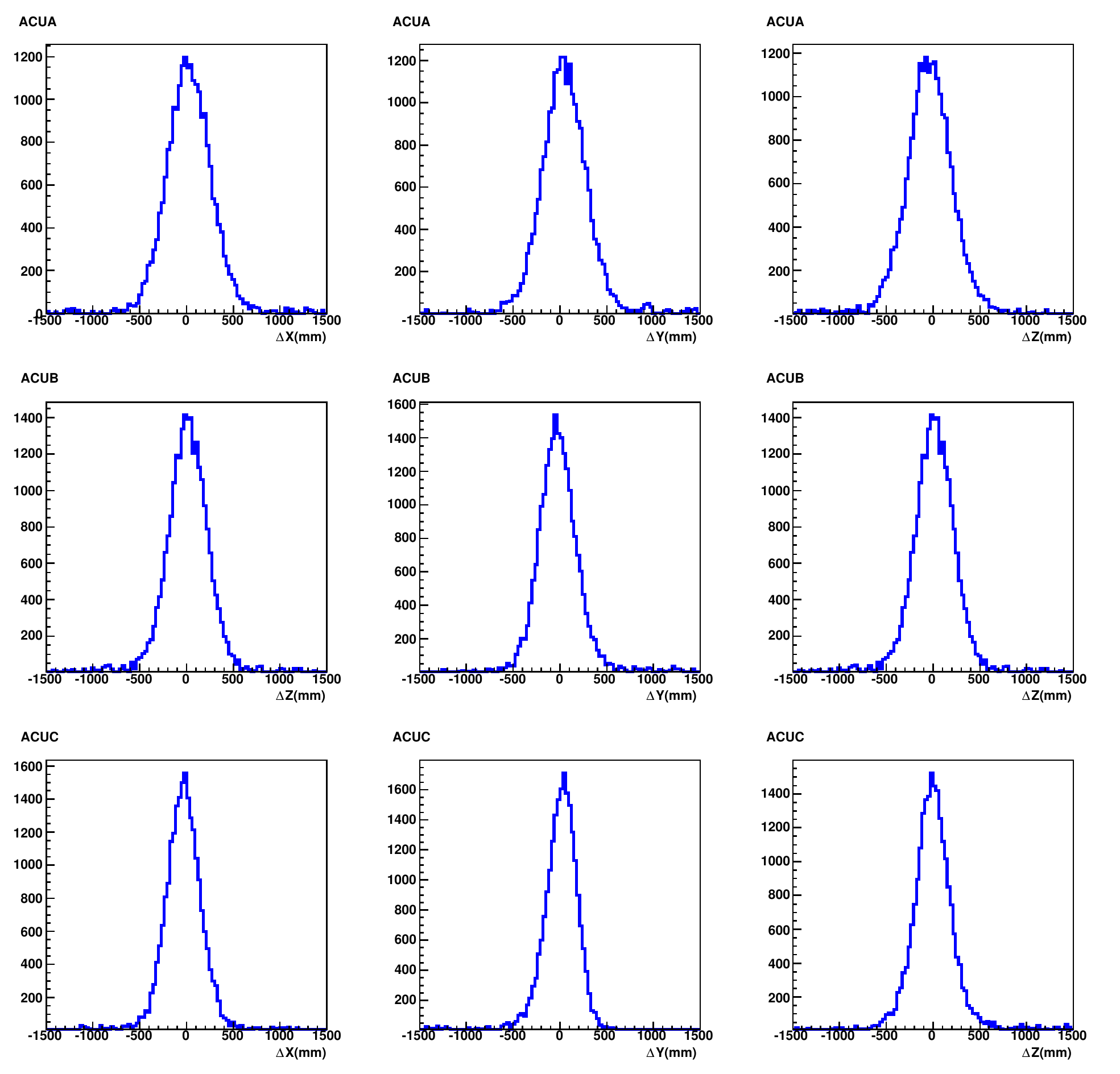}
\par\end{centering}
\caption{\label{fig:source-position-xyz}
  Difference in the reconstructed to true position of the $^{60}$Co source 
  deployed to $z$=0 for the three ACUs.
}
\end{figure}

Accurate knowledge of the true position of
the ACU sources provide stringent constraints to the vertex reconstruction
as well as vertex based energy correction. Fig.~\ref{fig:source-position}
shows the reconstructed position (with a charge pattern based algorithm) 
in $z$ and $r^2$ of $^{60}$Co source for all three ACUs at $z=0$ (center).
The projections of reconstructed to true position difference in $x$, $y$
and $z$ are shown separately for the three runs in 
Fig.~\ref{fig:source-position-xyz}.
Studies are ongoing to reduce the reconstruction bias and to improve 
the resolution.

\section{Summary}
\label{sec:summary}
In this paper we describe the design and performance of the automated 
calibration system for the Daya Bay Reactor Neutrino Experiment. 
This system consists of 24 functionally identical 
pulley/wheel and stepper motor assemblies, each 
designed and constructed as a precise and reliable robotic unit,
capable of deploying three difference sources vertically into detectors in 
a fully automated fashion. 
The Daya Bay Neutrino experiment has been in smooth full scale operation,
and has provided the first unambiguous measurement of neutrino
mixing angle $\sin^2(2\theta_{13})$. 
This fully automated system has provided a powerful and robust tool to 
monitor and calibrate the detector response at different positions in 
the detector, therefore giving a tight 
control on the uncertainty of the energy scale. 
In the remaining lifetime of the 
experiment we will continue using the system to help measure the neutrino 
rate and spectra shape from all eight detectors, targeting to a final 
precision <0.01 (90\% confidence level) to $\sin^2(2\theta_{13})$, and put 
tighter constraints on $\Delta m_{13}^2$ and the reactor neutrino spectrum. 

The calibration system reported in this paper is
uniquely designed within difficult size, material, automation, reliability, and 
interface constraints. The successful operation of this system provides 
precious experience to similar detectors where regular, remote, and 
automated calibration accesses are necessary.

%The Daya Bay Reactor Neutrino Experiment is designed to reach a sensitivity
%of <0.01 in $\sin^{2}2\theta_{13}$ and the first result has shown
%a great success of discovering a non-zero $\sin^{2}2\theta_{13}$
%at over 5-$\sigma$ standard deviations. Comparing with the competing
%experiments of the same generation, the success of the Daya Bay first
%result lies in two factors: sufficient statisics due to its unsurpassed
%target mass and the control of the dominant systematic uncertainties.
%The largest systematic uncertainty is the detection efficiency and
%its major component is from the energy scale calibration. The Daya
%Bay automatic calibration units have provided the most powerful tool
%and data to control the energy scale uncertainty. The successful design,
%installation and data analysis of the Daya Bay automatic calibration
%system can provide valuable experience to similar types of detectors.

\section{Acknowledgments}
This work was done with support from the US DoE, 
Office of Science, High Energy Physics, the US National Science Foundation, 
the Natural Science Foundation of China Grants 11005073 and 
11175116, the Shuguang Foundation of Shanghai Grant Z1127941, 
and Shanghai Laboratory for Particle Physics and Cosmology at 
the Shanghai Jiao Tong University. 

The authors gratefully acknowledge the
strong technical support of R. Cortez, J. Pendlay, and A. Raygoza of the 
Kellogg Radiation Laboratory at Caltech. This paper is dedicated to the 
memory of Ray Cortez, who had been a primary designer and 
chief machinist of this system. We acknowledge numbers of summer intern 
students from Caltech and Shanghai Jiao Tong University who 
had contributed to the design, 
fabrication, and installation of this system. We thank Xichao Ruan 
and his colleagues from Chinese Institute of Atomic Energy for providing 
radioactive sources to the system, and the DAQ and Slow Control group from the 
Institute of High Energy Physics for implementing interfaces to the 
calibration software. We are grateful to the rest of the Daya 
Bay collaborators, particularly Jeff Cherwinka 
from University of Wisconsin for his continuous professional 
engineering support throughout the years. We 
thank the Daya Bay on-site installation team for their dedicated work.



%% References


\bibliographystyle{model1-num-names}
\bibliography{<your-bib-database>}



\end{document}